\documentclass[journal]{IEEEtran}

\usepackage{amsmath}
\usepackage{amssymb}
\usepackage{cite}
\usepackage{graphicx}
\usepackage{url}
\usepackage{bbm}
\usepackage{stfloats}
\usepackage{algorithmic}
\usepackage{algorithm}
\usepackage{array}

\usepackage{txfonts}
\usepackage[thinlines]{easytable}

\newtheorem{proposition}{\textbf{Proposition}}
\newtheorem{lemma}{\textbf{Lemma}}

\newcommand\Pc{\ensuremath{{\cal P}}}

\DeclareMathOperator{\dB}{dB}

\DeclareMathOperator*{\st}{s.t.}

\newcommand\Ropt{\ensuremath{R_{1}^{\dag}}}

\graphicspath{{figure/}}
\begin{document}

\title{Short-Packet Downlink Transmission with Non-Orthogonal Multiple Access}
\author{Xiaofang Sun,~\IEEEmembership{Student Member,~IEEE,}
        Shihao Yan,~\IEEEmembership{Member,~IEEE,}
        Nan~Yang,~\IEEEmembership{Member,~IEEE,}\\
        Zhiguo Ding,~\IEEEmembership{Senior Member,~IEEE,}
        Chao Shen,~\IEEEmembership{Member,~IEEE,}
        and~Zhangdui Zhong,~\IEEEmembership{Senior Member,~IEEE}
\thanks{X. Sun, C. Shen, and Z. Zhong are with the State Key Lab of Rail Traffic Control and Safety, and the Beijing Engineering Research Center of High-speed Railway Broadband Mobile Communications, Beijing Jiaotong University, Beijing 100044, China (emails: \{xiaofangsun, chaoshen, zhdzhong\}@bjtu.edu.cn). S. Yan is with the School of Engineering, Macquarie University, Sydney, NSW, Australia (email: shihao.yan@mq.edu.au).  N. Yang is with the Research School of Engineering, Australian National University, Canberra, ACT 2601, Australia (email: nan.yang@anu.edu.au). Z. Ding is with the School of Electrical and Electronic Engineering, University of Manchester, Manchester, UK (email: Zhiguo.ding@gmail.com).}
\thanks{This paper has been presented in part at IEEE ICC workshop 2018~\cite{Conference}.}
}


\maketitle

\begin{abstract}
	
	This work introduces  downlink non-orthogonal multiple access (NOMA) into short-packet communications.  NOMA has great potential to improve fairness and spectral efficiency with respect to orthogonal multiple access (OMA) for low-latency downlink transmission, thus making it attractive for the emerging Internet of Things. We consider a two-user downlink NOMA system with finite blocklength constraints, in which the transmission rates and power allocation are optimized. To this end, we investigate the trade-off among the transmission rate, decoding error probability, and the transmission latency measured in blocklength.  Then, a one-dimensional search algorithm is proposed to resolve the challenges mainly due to the achievable rate affected by the finite blocklength and the unguaranteed successive interference cancellation. We also analyze the performance of OMA as a benchmark to fully demonstrate the benefit of NOMA. Our simulation results show that NOMA significantly outperforms OMA in terms of achieving a higher effective throughput subject to the same finite blocklength constraint, or incurring a lower latency to achieve the same effective throughput target. Interestingly, we further find that with the finite blocklength, the advantage of NOMA relative to OMA is more prominent when the effective throughput targets at the two users become more comparable. 
\end{abstract}


\section{Introduction}\label{sec:intro}

\subsection{Background and Motivation}

In the fifth generation (5G) wireless ecosystem, the majority of wireless connections will most likely be originated by autonomous machines and devices~\cite{Durisi_2016}. Against this background, machine-type communications (MTC) are emerging to constitute the basic communication paradigm in the Internet of Things (IoT)~\cite{NOMA_IoT}. The requirement on latency is stringent in some MTC scenarios~\cite{Durisi_2016},   e.g., intelligent transportation, factory automation (FA), and industry control systems,  since low latency is pivotal to ensure the real-time functionality in interactive communications of machines. Specifically, in some FA applications, the end-to-end latency at the application layer is required to be less than 1 ms~\cite{FA_2016}. This leads to a more strict  transmission latency requirement at the physical layer.

To support low-latency communications,  short-packet with finite blocklength codes is considered to reduce the transmission latency \cite{Khan_2017}. Specifically, in short-packet communications, as pointed out by~\cite{FBL_2010}, the decoding error probability at a receiver is not negligible since the blocklength is particularly small. This is different from Shannon's capacity theorem, in which the decoding error probability is negligible as the blocklength approaches infinity. Taking into account the effect of a finite blocklength, the achievable rate was examined in \cite{FBL_2010} to approximate the information-theoretic limit, which brings novelty in short-packet system design. This pioneering work serves as the foundation in examining the performance of short-packet communications. Triggered by~\cite{FBL_2010}, the impact of finite blocklength on different communication systems has been widely studied. For example,  the achievable transmission rate in quasi-static multiple-input multiple-output (MIMO) fading channels was examined in~\cite{Yang_2014}. The latency-critical packet scheduling was investigated in \cite{Xu_2016}. In terms of the channel coding schemes with finite blocklength, \cite{Bross_2010} and \cite{Lapidoth_2010} proposed optimal coding schemes in terms of the rate distortion function constrained to channel capacity under additive white Gaussian noise (AWGN) broadcast channel and medium access control (MAC) channel, respectively. In particular, these works showed that uncoded transmission schemes are optimal when the SNR is below some specific values. \cite{Persson_2012}  proposed a joint source-channel coding scheme  for a given finite blocklength and evaluated the performance of the scheme by mean square error distortion. \cite{Floor_2015} proposed a joint source-channel coding scheme with arbitrary blocklength on a Gaussian MAC channel and discussed  the impact of the blocklength on the performance of the proposed coding scheme.  

Furthermore, one challenge in MTC is the scalable and efficient connectivity for a massive number of devices sending short packets~\cite{Bockelmann_2016}. To address this challenge, different types of radio access technologies are investigated in the context of MTC~\cite{Osseiran_2014,NOMA_IoT_JSAC}. Particularly, non-orthogonal multiple access (NOMA) has attracted sharply increasing research interests as a promising technique for providing superior spectral efficiency~\cite{wong2017key}. The concept of downlink NOMA stems from the superposition coding on the degraded broadcast channel~\cite{NOMA_Degraged}, which has been widely studied in the literature from information theory perspective, e.g., \cite{Weingarten_2006,Romero_2017}. By leveraging NOMA, we can allocate more power to the users with poor channel qualities to ensure the achievable target rates at these users, thus striking a balance between network throughput and user fairness~\cite{Ding_2016_TWC}.  Specifically, NOMA can significantly increase the number of connected devices. This is due to the fact that NOMA allows for overloading spectrum by multiplexing users in the power domain~\cite{Ding_CM_2017}, which yields a higher flexibility and a more efficient use of spectrum and energy.   In order to unlock the benefit of NOMA, successive interference cancellation (SIC) is normally adopted at some users such that they can remove the co-channel interference caused by other users in NOMA and decode the desired signals successively~\cite{Cover_1972}.

Meanwhile, the benefit of NOMA has been widely examined in various wireless communications, such as broadcast channels~\cite{Sun_2016,Ding_2016_TWC}, full-duplex communications~\cite{Sun_2017_TCOM}, and physical layer security~\cite{PLS_2017}. Advocated by the unique benefit of multi-antenna systems, the application of MIMO techniques to NOMA was addressed in \cite{Ding_2016_TWC,Ding_TWC_2016}. Driven by the ever-increasing demand of high spectral and energy efficiencies, NOMA has also been applied to multi-cell networks~\cite{Han_MMTC_2014,Shin_COML_2016}. However, the potential benefit of NOMA in terms of latency reduction in the context of short-packet communications and the impact of finite blocklength on the performance of NOMA have rarely been examined. These leave an important gap in understanding on the benefit of NOMA in the context of short-packet communications and the impact of finite blocklength on NOMA, which motivate this work. Moreover, the number of users is generally large in IoT scenarios. To facilitate NOMA transmission, the users are scheduled to different clusters first. Then, each cluster can perform random access during an allowable time slot \cite{Niyato_2013}.  However, the clustering problem is an NP-hard problem. To reduce its complexity, some suboptimal solutions have been investigated, e.g.,   \cite{Liang_2017_matching,Liu_2016_CL}, but at the cost of system performance. Hence, it is pivotal to improve the spectral efficiency within a user cluster by resource allocation. This is also one of the motivations of this work.

\vspace{0mm}
\subsection{Our Main Contributions}
In this work, we introduce NOMA into short-packet communications and thoroughly examine its benefits in achieving a higher effective throughput relative to OMA subject to the same blocklength constraint, which in turn demonstrates the benefits of NOMA in latency reduction\footnote{The latency considered in this work is directly related to the blocklength.  The transmitter encodes the data bits to be delivered into the codeword with $N$ symbols.  Each symbol duration is defined as $T_s$. As such, the latency considered in this work is $NT_s$, which is proportional to the blocklength.}. We consider a specific MTC scenario, i.e., an FA scenario where an access point (AP) (e.g., a radio coordinator) has to transmit a certain amount of information to two stationary users (e.g., a process logic controller (PLC) and an actuator) within a short time period (requiring a low latency), enabling them to cooperatively perform some real-time functionalities. The two users are assumed to have been scheduled to one cluster and allocated to a resource block. We optimally design the transmission strategy for this cluster. Hence, this work serves as an important step for the further investigation of a massive-user scenario. To facilitate the optimal design,  we assume that the channel gains are available at the AP. To fully exploit the performance gain of NOMA over orthogonal multiple access (OMA), we assume that the channel gain disparity between the two users is large. Based on these assumptions, we examine how NOMA helps the AP communicating to the two users with a low latency, while keeping the OMA scheme as a benchmark. Different from the case of NOMA with infinite blocklength where perfect SIC can be always guaranteed \cite{Ding_CM_2017,Yang_2016_TWC}, the consideration of finite blocklength leads to the fact that the perfect SIC may not be guaranteed. This brings about new challenges in the optimal design of the NOMA scheme, which have been addressed in this work. To take into account the impact of the non-zero decoding error probability caused by finite blocklength, we adopt the effective throughput as the metric to evaluate the system performance.

Different from our previous work \cite{Conference},  we provide insights and analysis on the constraints and determine the optimal solution to the optimization problem in this work. Moreover, we propose a fixed-point iteration algorithm which allows us to seek the optimal transmission rate for the weak channel user with a higher computational efficiency than one-dimensional search in this work, but not \cite{Conference}. The main contributions of this work are summarized as below.

\begin{itemize}
	\item We explicitly determine the optimal design of the NOMA scheme, in which the transmission rates and power allocation are optimized. To strike a balance between the system throughput and user fairness, this optimization maximizes the effective throughput of the user with a higher channel gain while guaranteeing a certain effective throughput target at the other user. To address the challenges caused by the complex capacity formula and unguaranteed SIC, we first analytically prove that the equality in the power constraint is active and the effective throughput target imposed at the user with a lower channel gain is always ensured. Then, we detail the steps to achieve the optimal transmission rates and power allocation. Besides, a computationally efficient algorithm, i.e.,  fixed-point iteration algorithm, is proposed for seeking the optimal transmission rate for the weak channel user.

	\item In order to explicitly demonstrate the benefit of NOMA in the context of short-packet communications, we take the optimal design of the OMA scheme as a benchmark, in which the optimal time slot allocation has to be determined on top of the optimal transmission rates and power allocation. 
	
	\item Considering practical application scenarios with a finite blocklength, a thorough comparison between the NOMA and OMA schemes is provided. Our examination indicates that the NOMA scheme can significantly outperform the OMA scheme in terms of achieving a higher effective throughput at one user (subject to the same constraint on the effective throughput at the other user) with the same latency or incurring a lower latency to achieve the same effective throughput targets. Interestingly, the advantage of NOMA relative to OMA is more dominant when the effective throughput targets at the two users become more comparable, which is different from their comparison result  with an infinite blocklength.
\end{itemize}

The rest of the paper is organized as follows. Section \ref{sec:System} presents the system model and formulates the optimization problem. Section \ref{sec:Transmission_NOMA} details the transmission strategy in the NOMA scheme. In Section \ref{sec:NOMA}, the optimal design of the NOMA scheme is provided. Section \ref{sec:OMA_opt} presents the optimal design of the OMA scheme.  Numerical results are presented in Section \ref{sec:simulation} to draw useful insights. Finally, Section \ref{sec:conclusion} concludes this work.

\vspace{-0ex}
\section{System Model and Problem Formulation}\label{sec:System}
\subsection{System Model}
\begin{figure}[!t]
	\centering
	\includegraphics[width=3in]{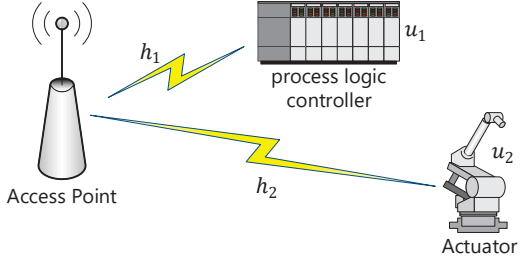}\vspace{0mm}
	\caption{System model of interest, where a single-antenna AP communicates with two single-antenna users with different channel gains.}\label{fig:system_model}\vspace{0mm}
\end{figure}
In this work, we consider a downlink broadcast FA scenario as depicted in Fig.~\ref{fig:system_model}, in which a single-antenna AP (e.g., a radio coordinator) serves two stationary single-antenna users (e.g., a PLC and an actuator) within a finite blocklength of $N$ symbol periods. As the AP is equipped with a single antenna, it serves the two users either over orthogonal resource blocks with the OMA scheme, or over the same resource block with the NOMA scheme. In this work, we mainly focus on the NOMA scheme, while the OMA scheme serves as a benchmark. In the NOMA scheme, the two users are assumed to have been scheduled to one cluster and  allocated to a resource block. Since the users' locations are fixed, the channel gains vary slowly in time.  We assume that the channel gains from the AP to the users are available at the AP and users\footnote{To estimate the channel gains,   we follow \cite{NOMA_IoT_JSAC} and assume that the users can perform the channel estimation by using regular pilot signals transmitted by the AP.  The channel estimation method proposed in  \cite{NOMA_IoT_JSAC} is particularly advantageous for fixed-location applications.}. To fully exploit the benefit of NOMA over OMA, we assume that the channel gains of the two users are significantly different. Without loss of generality, we define the user with the higher channel gain (the norm of the channel is higher) as user 1 (denoted by u$_1$) and the other one as user 2 (denoted by u$_2$). The channel coefficients from the AP to u$_1$ and from the AP to u$_2$ are denoted by $\tilde{h}_1$ and $\tilde{h}_2$, respectively. We assume that $\tilde{h}_1$ and $\tilde{h}_2$ are subject to independent quasi-static Rayleigh fading with equal blocklength $N$. Given the channel gain relationship between u$_1$ and u$_2$, we have $|\tilde{h}_1|> |\tilde{h}_2|$.

\vspace{-1ex}
\subsection{Achievable Transmission Rate with Finite Blocklength}

As per Shannon's coding theorem, the decoding error probability at the receiver becomes negligible as the blocklength approaches infinity \cite{Shannon_1948}. Differently, short-packet communication aims to achieve low latency (e.g., short delay), in which the blocklength needs to be finite and typically small. As pointed out by \cite{FBL_2010}, the decoding error probability at the receiver is non-negligible when the blocklength is finite. Specifically, perfect SIC cannot be guaranteed at the receiver side in this work, due to the finite blocklength constraint.  Taking into account the impact of the non-zero error probabilities on successively decoding the signals at the receivers, we introduce the effective error probability,  which is defined by  $\overline{\epsilon}_i, i\in\{1,2\}$, to denote the actual error probability at u$_i$. Furthermore, considering the trade-off between error probability and transmission rate, we adopt the effective throughput as the metric to evaluate the system performance with finite blocklength. Mathematically, the effective throughput at u$_i$ is defined by 
\begin{align}\label{eq:Throughput}
\overline{T}_i = \frac{N_i}{N} R_i (1 - \overline{\epsilon}_i)~(\rm bps/Hz),
\end{align}
where  $N_i$ is blocklength allocated to u$_i$ and $R_i$ is the transmission rate in the finite blocklength regime  at u$_i$. 

Based on \cite{FBL_2010,Ozcan_2013}, with finite blocklength $N_i$ for a given decoding error probability $\epsilon_i$ at user $i$, $i\in\{1,2\}$, $R_i$ can be accurately approximated by
\begin{align}\label{eq:Rate_FBL}
R_i \approx\log_2(1+\gamma_i)-\sqrt{\frac{V_i}{N_i}}\frac{Q^{-1}(\epsilon_i)}{\ln 2},
\end{align}
where $\gamma_i$ denotes the received signal-to-noise ratio (SNR) at user $i$, $Q^{-1}(\cdot)$ is the inverse function of $Q(x)=\int_{x}^{\infty}\frac{1}{\sqrt{2\pi}}\exp(-\frac{t^2}{2})dt$, and $V_i$ is the channel dispersion. Specifically, based on the results in \cite{Yang_2014}, the expression for $V_i$ for the single-antenna quasi-static Rayleigh fading channel is  $V_i=1-(1+\gamma_i)^{-2}$.

Notably, the authors in \cite{FBL_2010} have compared the performance of a certain family of multiedge LDPC codes decoded via a low-complexity belief-propagation decoder against this finite blocklength fundamental limit. The simulation results showed that the relative gap to the finite blocklength fundamental limit is approximately constant. Furthermore, LDPC, Reed-Muller, Polar, and BCH codes have been investigated  in \cite{Li_2012_COML,Wonterghem_2016}.  The aforementioned studies have verified \eqref{eq:Rate_FBL}. Therefore, we adopt \eqref{eq:Rate_FBL} as a preliminary result for our work.

For a given transmission rate $R_i$, the decoding error probability at user $i$ is approximated by
\begin{align}\label{eq:EP}
\epsilon_i  \approx  Q\left(f(\gamma_i,N_i,R_i)\right),
\end{align}
where $f(\gamma_i,N_i,R_i) \triangleq \ln2\sqrt{\frac{N_i}{1-(1_+\gamma_i)^{-2}}}\left(\log_2\left(1+\gamma_i\right)-R_i\right)$, and  $0\leq \epsilon_i\leq 0.5$ is for the sake of practical reliable communication.



\subsection{Optimization Problem}
In the considered system, the AP needs to serve u$_1$ and u$_2$ within $N$ symbol periods (i.e., the finite blocklength is $N$). In addition, the AP fully consumes the symbol periods to transmit signals for ensuring the reliability.  The ultimate goal is to achieve the maximum effective throughput at u$_1$, while guaranteeing a specific constraint on the effective throughput at u$_2$ subject to a total power constraint. Mathematically, the optimization problem for the AP is formulated as
\begin{subequations}\label{eq:Opt}
	\begin{align}
	\underset{\Delta}{\max} ~ &\overline{T}_1 \label{eq:Ojective}\\
	\st~&\overline{T}_2 \geq T_0,\label{eq:C_T}\\
	& P_1N_1+P_2N_2\leq PN,\label{eq:C_Energy}\\
	&(N_1,N_2)\in \begin{cases}
	\{(\mathcal{N}_1,\mathcal{N}_2)|\mathcal{N}_1=\mathcal{N}_2 = N\}, &\hspace{-1mm}\text{for NOMA,}\\
	\{(\mathcal{N}_1,\mathcal{N}_2)|\mathcal{N}_1+\mathcal{N}_2=N\}, &\hspace{-1mm}\text{for OMA,}
	\end{cases}\label{eq:C_time}
	\end{align}
\end{subequations}
where $\Delta=\{R_1, R_2, P_1, P_2, N_1, N_2\}$ represents the variable set that needs to be determined at the AP. It is noted that based on the expressions for the transmission rate and error probability in \eqref{eq:Rate_FBL} and \eqref{eq:EP}, respectively,   we find that the transmission rate, error probability, and the allocated power are coupled together.  Given any two of them, the third one is uniquely determined. This indicates that any two of them can be viewed as optimization variables. We further note that the transmission rates and power allocation are determined by the AP. Thus, we adopt the two variables to facilitate the system design and control the decoding error probabilities at the receiver side. $T_{0}$ is the minimum required effective throughput at u$_2$, $P_1$ and $P_2$ are the allocated transmit powers to u$_1$ and u$_2$, respectively, $P$ is the average transmit power within one fading block. \eqref{eq:C_time} imposes the finite blocklength on the two users.   As per \eqref{eq:Opt}, we need to design the symbol period allocation at the AP on top of determining the power allocation and transmission rates for u$_1$ and u$_2$, such that $\overline{T}_1$ is maximized subject to the  given constraints.

\section{Transmission Strategies with NOMA}\label{sec:Transmission_NOMA}

In this section, we focus on the design of NOMA transmission. We first specify the optimization problem given in \eqref{eq:Opt} for NOMA. Then, we detail the transmissions to user 1 and user 2 and mathematically characterize the optimization problem associated with the NOMA transmission.

\subsection{Optimization Problem in NOMA}

In NOMA, superposition coding (SC) is employed in the transmission such that the AP is able to transmit signals to u$_1$ and u$_2$ simultaneously at different power levels. As such, we have $N_1 = N_2=N$ in the NOMA transmission strategy. Then, when NOMA is adopted at the AP, the optimization problem given in \eqref{eq:Opt} yields
\begin{subequations}\label{eq:Opt_NOMA}
	\begin{align}
	\underset{\Delta_n}{\max} ~~~&\overline{T}_1 \label{eq:Objective_NOMA}\\
	\st~~~~&P_1+P_2 \leq P, \label{eq:C_Energy_NOMA}\\
	&\overline{T}_2\geq T_0, \label{eq:C_T_NOMA}
	\end{align}
\end{subequations}
where $\Delta_n=\{R_1, R_2, P_1, P_2\}$ is the variable set that needs to be determined at AP for NOMA transmission. To facilitate the optimal design, we then detail the NOMA transmission strategy and derive the expressions for $\overline{T}_1$ and $\overline{T}_2$ in the following two subsections, respectively.

\subsection{Transmission to User 1}

In the NOMA transmission, the transmitted signal at the AP is given by
\begin{align}\label{eq:x}
x=\sqrt{P_1}x_1+\sqrt{P_2}x_2,
\end{align}
where $x_1\sim\mathcal{CN}(0,1)$ and $x_2\sim\mathcal{CN}(0,1)$ represent the information bearing signals to u$_1$ and u$_2$, respectively, following circularly symmetric complex Gaussian (CSCG) distribution with zero mean and variance one. It is assumed that $x_1$ and $x_2$ are independent and identically distributed.


Then, the received signal at u$_1$ at each symbol period is given by
\begin{align}\label{eq:y1}
y_1 = \tilde{h}_1x+n_1 = \tilde{h}_1 (\sqrt{P_1}x_1+\sqrt{P_2}x_2)+n_1,
\end{align}
where $n_1\sim \mathcal{CN}(0,\sigma_1^2)$ denotes the AWGN at u$_1$ following CSCG  distribution with zero mean and variance $\sigma_1^2$. Due to $|\tilde{h}_1|>|\tilde{h}_2|$, we consider that SIC is employed at u$_1$ to remove the interference caused by $x_2$. To this end, u$_1$ first decodes $x_2$ while the interference caused by $x_1$ is treated as noise (e.g.,~\cite{Ding_CM_2017}). Following \eqref{eq:y1}, the signal-to-interference-plus-noise ratio (SINR) of $x_2$ at u$_1$, denoted by $\gamma_2^1$, is given by
\begin{align}\label{eq:SINR_SIC}
\gamma_2^1=\frac{P_2h_1}{P_1h_1+1},
\end{align}
where $h_1=\frac{\left|\tilde{h}_1\right|^2}{\sigma_1^2}$ denotes the normalized channel gain from AP to u$_1$.

For the sake of clarity, we denote $\mathbb{D}_i^j=0$ as the event that $x_i$ is correctly decoded at u$_j$, $i,j \in \{1,2\}$, while denote $\mathbb{D}_i^j=1$ as the event that $x_i$ is incorrectly decoded at u$_j$. The probability of an event occurring is denoted by $\Pc(\cdot)$.
Following \eqref{eq:EP}, the decoding error probability of $x_2$ at u$_1$ (the outage probability of SIC) for a given $R_2$, denoted by $\epsilon_2^1$, is approximated by
\begin{align}\label{eq:EP21}
\epsilon_2^1=\Pc(\mathbb{D}_2^1=1)=Q\left(f(\gamma_2^1, N_2, R_2)\right).
\end{align}
Accordingly, the probability that $x_2$ is correctly decoded and completely canceled at u$_1$ is $1-\epsilon_2^1$. This indicates that perfect SIC may not be guaranteed in NOMA with finite blocklength, differing from that in NOMA with infinite blocklength where the perfect SIC can always be guaranteed~\cite{Ding_CM_2017}. Taking into account the impact of the non-zero error probabilities for successively decoding the signals at u$_1$, the effective decoding error probability of $x_1$ at u$_1$ is achieved by the marginal probability, which is given by
\begin{align}\label{eq:effective_error_1}
\overline{\epsilon}_1 = \Pc(\mathbb{D}_1^1=1)=\sum_{i=0}^1\Pc(\mathbb{D}_1^1=1|\mathbb{D}_2^1=i)\Pc(\mathbb{D}_2^1=i).
\end{align}
To address this issue, we next derive $\Pc(\mathbb{D}_1^1=1|\mathbb{D}_2^1=0)$ and $\Pc(\mathbb{D}_1^1=1|\mathbb{D}_2^1=1)$ for $\overline{\epsilon}_1$.

If SIC succeeds, i.e., $\mathbb{D}_2^1=0$, (which occurs with the probability $1-\epsilon_{2}^{1}$), following \eqref{eq:y1} the SNR of $x_1$ at u$_1$, denoted by $\gamma_1$, is given by
\begin{align}\label{eq:SNR1}
\gamma_1=P_1h_1.
\end{align}
Accordingly, the decoding error probability of $x_1$ at u$_1$ for a given $R_1$ conditioned on the guaranteed SIC (i.e., $\mathbb{D}_2^1=0$), denoted by $\epsilon_1$, is approximated by
\begin{align}\label{eq:EP1}
\epsilon_1 =\Pc(\mathbb{D}_1^1=1|\mathbb{D}_2^1=0)=Q\left(f(\gamma_1,N_1,R_1)\right).
\end{align}
In this case, the throughput at u$_1$, denoted by $T_1$, is given by $T_1=R_1\left(1-\epsilon_1\right)$.

Alternately, if SIC fails, i.e., $\mathbb{D}_2^1=1$, (which occurs with the probability $\epsilon_{2}^{1}$), u$_1$ has to decode $x_1$ directly subject to the interference caused by $x_2$. Correspondingly, following \eqref{eq:y1} the SINR of $x_1$ at u$_1$, denoted by $\gamma'_{1}$, is given by
\begin{align}\label{eq:SINR1}
\gamma'_{1}=\frac{P_1h_1}{P_2h_1+1}.
\end{align}
In general, $\gamma'_{1}$ is significantly less than $\gamma_1$ given in \eqref{eq:SNR1}, since more power is allocated to u$_2$. As such, we may have $R_1 > \log_2(1 + \gamma'_{1})$ in the design of NOMA. Considering the case with $R_1 > \log_2(1 + \gamma'_{1})$, the decoding error probability of $x_1$ for a given $R_1$ conditioned on the failed SIC (i.e., $\mathbb{D}_2^1=1$), denoted by $\epsilon'_{1}$, is approximated by
\begin{align}\label{eq:EP1'}
\epsilon'_{1} &= \Pc(\mathbb{D}_1^1=1|\mathbb{D}_2^1=1)\notag\\
&=\begin{cases}
Q\left(f(\gamma'_{1}, N_1, R_1)\right),&\text{if}~~ R_1\leq \log_2(1+\gamma_1'),\\
1,&\text{if}~~R_1 > \log_2(1+\gamma_1').
\end{cases}
\end{align}
Then, the throughput at u$_1$, denoted by $T'_{1}$, is given by $T'_{1}=R_1 (1- \epsilon'_{1})$.

Following \eqref{eq:effective_error_1}, we obtain the effective decoding error probability of $x_1$ at u$_1$ as
\begin{align}\label{eq:Average_EP1}
\overline{\epsilon}_1=\epsilon_1-\epsilon_1\epsilon_{2}^{1}+\epsilon_{2}^{1}\epsilon'_{1}.
\end{align}
As such, the effective throughput at u$_1$ is given by
\begin{align}\label{eq:Average_T1}
\overline{T}_1&=R_1(1-\epsilon_1+\epsilon_1\epsilon_{2}^{1}-\epsilon_{2}^{1}\epsilon'_{1}).
\end{align}


\subsection{Transmission to User 2}

Following \eqref{eq:x}, the received signal at u$_2$, when AP adopts NOMA, is given by
\begin{align}\label{eq:y2}
y_2=\tilde{h}_2x+n_2=\tilde{h}_2(\sqrt{P_1}x_1+\sqrt{P_2}x_2)+n_2,
\end{align}
where $n_2\sim\mathcal{CN}(0,\sigma_{2}^2)$ denotes the AWGN at u$_2$ with zero mean and variance $\sigma_{2}^2$.

Due to $|\tilde{h}_{1}| > |\tilde{h}_{2}|$, SIC is not conducted at u$_2$. As such, u$_2$ decodes its own signal directly subject to the interference caused by $x_1$. Following \eqref{eq:y2}, the SINR of $x_2$ at u$_2$, denoted by $\gamma_2$, is given by
\begin{align}\label{eq:SINR2}
\gamma_2=\frac{P_2h_2}{P_1h_2+1},
\end{align}
where $h_2=\frac{|\tilde{h}_2|^2}{\sigma_2^2}$ denotes the normalized channel gain from AP to u$_2$.
Accordingly, the decoding error probability of $x_{2}$ at u$_2$ for given $R_{2}$, denoted by $\epsilon_2$, is approximated by
\begin{align}\label{eq:EP2}
\epsilon_2 = \Pc(\mathbb{D}_2^2=1)= Q\left(f(\gamma_2, N_2, R_2)\right).
\end{align}

Since there only exists one decoding strategy at u$_2$, the decoding error probability $\epsilon_2$ is actually the effective decoding error probability at u$_2$, i.e., $\overline{\epsilon}_2=\epsilon_2$.
Then, the effective throughput at u$_2$, $\overline{T}_2$, is given by
\begin{align}\label{eq:T2}
\overline{T}_2=R_2(1-\overline{\epsilon}_2)=R_2(1-\epsilon_2).
\end{align}

With the expressions for $\overline{T}_1$ and $\overline{T}_2$ given in \eqref{eq:Average_T1} and \eqref{eq:T2}, respectively, the optimization problem given in \eqref{eq:Opt_NOMA} is well defined. We next tackle this optimization problem in the following section.

\section{Design of Transmission Rates and Power Allocation in NOMA}\label{sec:NOMA}

In this section, we focus on the design of the transmission rates and power allocation in the NOMA transmission, i.e., focus on solving the optimization problem given in \eqref{eq:Opt_NOMA}. To this end, we first provide analysis and insights on the constraints and then find the optimal solution.

\subsection{Equalities in Constraints}\label{subsec:equal_constraint}

In this subsection, we tackle the two constraints given in \eqref{eq:C_Energy_NOMA} and \eqref{eq:C_T_NOMA} in order to facilitate solving the optimization problem given in \eqref{eq:Opt_NOMA}.

For the sake of clarity, in the NOMA scheme, we replace both $N_1$ and $N_2$ with $N$, and we substitute $f(\gamma_i,R_i)$ for $f(\gamma_i,N_i,R_i)$ in \eqref{eq:EP},  since we have $N_1=N_2=N$ in NOMA.  In order to simplify constraint given in \eqref{eq:C_Energy_NOMA}, we first examine the monotonicity of the error probability $\epsilon_i$ given in \eqref{eq:EP} with respect to the corresponding SNR/SINR $\gamma_i$ in the following theorem.
\begin{proposition}\label{lemmaEP_P}
	The decoding error probability given in \eqref{eq:EP} is a monotonically decreasing function of the corresponding SNR/SINR.
	\begin{IEEEproof}
		The detailed proof is provided in Appendix~\ref{App:EP_Gamma}.
	\end{IEEEproof}
\end{proposition}

We note that in Proposition~\ref{lemmaEP_P}, the decoding error probability can be any one of $\epsilon_1$, $\epsilon_1'$, $\epsilon_2^1$, and $\epsilon_2$. Based on Proposition~\ref{lemmaEP_P}, we next prove in the following lemma that the equality in \eqref{eq:C_Energy_NOMA} is always guaranteed.
\begin{lemma}\label{Lemma_P}
	The equality in the power constraint \eqref{eq:C_Energy_NOMA}, i.e., $P_1+P_2=P$, is always guaranteed in order to maximize $\overline{T}_1$ subject to $\overline{T}_2\geq T_0$.
	\begin{IEEEproof}
		The detailed proof is provided in Appendix~\ref{App:Power}.
	\end{IEEEproof}
\end{lemma}

Lemma~\ref{Lemma_P} indicates that the AP fully consumes the maximum transmit power to maximize $\overline{T}_1$ subject to $\overline{T}_2\geq T_0$. This lemma significantly facilitates the power allocation at the AP, since it shows that as long as we can determine the power allocation to one user, all the remaining power needs to be allocated to the other user.

In the following lemma, we prove that the equality in the constraint~\eqref{eq:C_T_NOMA} is also always guaranteed.
\begin{lemma}\label{Lemma_T2}
	The equality in the effective throughput constraint \eqref{eq:C_T_NOMA} is always guaranteed, i.e., $\overline{T}_2=T_0$, in order to maximize $\overline{T}_1$ subject to $\overline{T}_2\geq T_0$.
	\begin{IEEEproof}
		To facilitate the proof, we first exploit the monotonicity of the decoding error probability given in \eqref{eq:EP} with respect to $R_i$. To this end, the partial derivative of $\epsilon_i$ with respect to $R_i$ is given by
		\begin{align}\label{prooft2}
		\frac{\partial \epsilon_i}{\partial R_i}=\frac{\sqrt{N}\ln 2}{\sqrt{2\pi}\sqrt{1-(1+\gamma_i)^{-2}}}e^{-\frac{f^2\left(\gamma_i,R_i\right)}{2}},
		\end{align}
		which is always larger than zero. Accordingly, we have that $\epsilon_i$ is a monotonically increasing function of $R_i$. 
		
		We next prove by contradiction that the equality in \eqref{eq:C_T_NOMA} is active at the optimal solution. We first suppose that in the optimal solution we have $\overline{T}_2>T_0$, where the maximum value of $\overline{T}_1$ is $T_1^{\dag}$.
		Then, we can reduce $\overline{T}_2$ to $T_0$ by using a smaller $R_2$ while keeping the power allocation fixed, since $\overline{T}_2$ is a continuous function of $R_2$ and $\overline{T}_2 = 0$ when $R_2 = 0$. By doing so, the outage probability of SIC, i.e., $\epsilon_2^1$, decreases, since $\epsilon_2^1$ monotonically increases with $R_2$ as per \eqref{prooft2}. As we proved in Appendix~\ref{App:Power} that $\overline{\epsilon}_1$ monotonically increases with $\epsilon_{2}^{1}$,  $\overline{T}_1$ increases as $R_2$ decreases. This contradicts to the claim of optimality  that $T_1^{\dag}$ is the maximum value of $\overline{T}_1$. Therefore, the equality in the effective throughput constraint must hold for the optimal design.
	\end{IEEEproof}
\end{lemma}

Based on this lemma, the constraint $\overline{T}_2=T_0$ uniquely determines the one-to-one relationship between the transmit power $P_2$ and the corresponding optimal transmission rate $R_2$, which will be discussed in the next subsection.

\subsection{Optimal Transmission Design}\label{subsec:Optimal_Design}

Following the aforementioned analysis, in this subsection we determine the optimal solution to the optimization problem given in \eqref{eq:Opt_NOMA}.

Based on \eqref{eq:EP21}, the perfect SIC cannot be always guaranteed in the NOMA scheme with a finite blocklength. As such, taking into account the outage probability of SIC, i.e., $\epsilon_2^1$, leads to a fact that the objective function $\overline{T}_1$ does not always increase with $P_1$ subject to $P_1+P_2=P$. As $\epsilon_2^1$ increases with $P_1$ based on Proposition~\ref{lemmaEP_P}. But on the other hand, the decoding error probabilities $\epsilon_1$ and $\epsilon_1'$ at u$_1$ decrease with $P_1$. Accordingly, there is a non-trivial trade-off between the effective throughput $\overline{T}_1$ and $P_1$. This fact is different from the NOMA scheme with an infinite blocklength, where $\overline{T}_1$ monotonically increases with $P_1$ subject to $P_1+P_2=P$. This fact also brings in challenges in the design of the optimal power allocation and transmission rates in the NOMA scheme with a finite blocklength.

To address the previous issue, we next detail the main steps to determine the optimal design of the power allocation and transmission rates in the NOMA scheme.

\textbf{Step~1}: Determine $R_2$ for a feasible $P_2$.

As $x_2$ is directly decoded at u$_2$ by treating $x_1$ as noise, for a given power allocation the effective throughput achieved at u$_2$ is independent of $R_1$.  Inspired by this, we first determine the value of $R_2$ that maximizes $\overline{T}_1$ for given feasible $P_1$ and $P_2$. The feasibility of $P_2$ will be discussed in Step 3.

We define the effective throughput $\overline{T}_2$ in terms of $R_2$ by $\mathcal{T}(R_2)\triangleq R_2(1-Q(f(\gamma_2,R_2)))$. To facilitate the design of $R_2$, we examine the monotonicity and concavity of $\mathcal{T}(R_2)$ with respect to $R_2$ in the following lemma.

\begin{lemma}\label{Lemma_T2_R2}
	$\mathcal{T}(R_2)$ does not monotonically increase with $R_2$ but is concave with respect to $R_2$.
	\begin{IEEEproof}
		The detailed proof is provided in Appendix~\ref{APP:T2_R2}.
	\end{IEEEproof}
\end{lemma}

Following Lemma~\ref{Lemma_T2}, we note that for a feasible $P_2$, the optimal value of $R_2$ satisfies $\mathcal{T}(R_2)=T_0$. Lemma~\ref{Lemma_T2_R2} indicates that for a given feasible $P_2$,  there are two values of $R_2$ that satisfy  $\overline{T}_2=T_0$. The smaller one that satisfies $\mathcal{T}'(R_2)\geq 0$, denoted by $R_2^{\dag}$, is the optimal value that maximizes $\overline{T}_1$. This is due to the fact that $\overline{T}_1$ monotonically decreases with $R_2$, which is proved in Lemma~\ref{Lemma_P}. However, $R_2$ is an input argument of $Q$-function for $\overline{T}_2$ given by  \eqref{eq:T2}. This prevents us from deriving a closed-form expression for $R_2$. To address this issue,  a fixed-point iteration algorithm \cite{Lipschitz} is proposed in the following proposition for seeking $R_2^{\dag}$.

\begin{proposition}\label{Propostion:R2}
	The solution of $R_2$ to  $\mathcal{T}(R_2)=T_0$ can be obtained by the fixed-point iteration
	\begin{align}\label{eq:FixedPointIteration}
	R_2:=\mathcal{F}(R_2)=\frac{T_0}{1-Q\left(f(\gamma_2,R_2)\right)}.
	\end{align}
	\begin{IEEEproof}
		Based on the Theorem 2.1 in \cite{Lipschitz}, the iteration converges to a fixed point if $\max\{\left|\mathcal{F}'(R_2) \right|\}< 1$ is guaranteed for $R_2$ that satisfies $\mathcal{T}'(R_2)\geq 0$, i.e.,  $0\leq R_2\leq R_2^{\ddag}$, where $R_2^{\ddag}$ denotes the value that satisfies $\mathcal{T}'(R_2^{\ddag}) = 0$.  Notably, $\mathcal{T}'(R_2)\geq 0$ guarantees that the iteration converges to the smaller value that satisfies  $\mathcal{T}(R_2)=T_0$.  To verify the convergence of the iteration, the first derivative of $\mathcal{F}(R_2)$ with respect to $R_2$ is derived as 
		\begin{align}\label{eq:FixedPointFirstOder}
		\mathcal{F}'(R_2)=\frac{T_0}{\sqrt{2\pi}b}e^{-\frac{f^2(\gamma_2,R_2)}{2}}\left(1-Q(f(\gamma_2,R_2))\right)^{-2}\geq 0,
		\end{align}
		where $b=\frac{\sqrt{N}\ln 2}{\sqrt{1-(1+\gamma_2)^{-2}}}$. 	$\mathcal{F}'(R_2)$ monotonically increases with $R_2$, as $\mathcal{F}''(R_2)\geq 0$. The proof of $\mathcal{F}''(R_2)\geq 0$ is omitted here due to  page limit.
		
		Then,  we have $0\leq \mathcal{F}'(R_2)\leq \mathcal{F}'(R_2^{\ddag})$ for $0\leq R_2\leq R_2^{\ddag}$. Based on \eqref{eq:f'R2}, we have $\frac{1}{\sqrt{2\pi}b}e^{-\frac{f^2(\gamma_2,R_2^{\ddag})}{2}}=\frac{1-Q(f(\gamma_2,R_2^{\ddag}))}{R_2^{\ddag}}$. Substituting $\frac{1-Q(f(\gamma_2,R_2^{\ddag}))}{R_2^{\ddag}}$ for $\frac{1}{\sqrt{2\pi}b}e^{-\frac{f^2(\gamma_2,R_2^{\ddag})}{2}}$ in \eqref{eq:FixedPointFirstOder}, we have
		\begin{align}
		\mathcal{F}'(R_2^{\ddag})=\frac{T_0}{R_2^{\ddag}\left(1-Q(f(\gamma_2,R_2^{\ddag}))\right)}.
		\end{align}
		Notably, for any feasible $P_1$ and $P_2$, to ensure the effective throughput target can be achieved at u$_2$ based on Step 3,  $R_2^{\ddag}\left(1-Q(f(\gamma_2,R_2^{\ddag}))\right)\geq T_0$ is guaranteed. This indicates that $\mathcal{F}'(R_2^{\ddag})< 1$, otherwise $R_2^\ddag$ is the solution to $\mathcal{T}(R_2)=T_0$.
		
		Therefore, we prove that $\max\{\left|\mathcal{F}'(R_2) \right|\}< 1$ holds for  $R_2$ that satisfies $\mathcal{T}'(R_2)\geq 0$.

	\end{IEEEproof}
	
\end{proposition}

\textbf{Step~2}: Determine $R_1$ for given $P_1$, $P_2$, and $R_2$.

Following the previous optimal design for $R_2$, we then determine the value of $R_1$ that maximizes $\overline{T}_1$ for given $P_1$, $P_2$, and $R_2$ in the following theorem.
\begin{proposition}\label{Proposition_R1}
	The value of $R_1$ that maximizes the effective throughput $\overline{T}_1$ for given $P_1$, $P_2$, and $R_2$ is given by
	\begin{align}\label{eq:R1*}
	\Ropt = \begin{cases}
	\overline{R}_1^{\dag}, &\text{if}~~0\leq R_1\leq \log_2(1+\gamma_1'),\\
	\widetilde{R}_1^{\dag}, &\text{if}~~\log_2(1+\gamma_1')< R_1 \leq \log_2(1+\gamma_1),\\
	\end{cases}
	\end{align}
	where $\overline{R}_1^{\dag}$ is the unique solution to $(1-\epsilon_2^1)\mathcal{U}(\gamma_1,\overline{R}_1^{\dag})+\epsilon_2^1\mathcal{U}(\gamma_1',\overline{R}_1^{\dag})=0$. 
	$\widetilde{R}_1^{\dag}$ is the unique solution to $\mathcal{U}(\gamma_1,\widetilde{R}_1^{\dag})=0$.
	And function $\mathcal{U}(x,y)$ is defined by
	\begin{align}\label{eq:First_order_T}
	\mathcal{U}(x,y)\triangleq 1-Q\left(f(x, y)\right)+\frac{y \frac{\partial f(x,y)}{\partial y}}{\sqrt{2\pi}}e^{-\frac{f^2(x,y)}{2}},
	\end{align}
	where $ \frac{\partial f(x,y)}{\partial y}=-{N\ln 2}/{\sqrt{1-(1+x)^{-2}}}$.
	\begin{IEEEproof}
		The detailed proof is provided in Appendix~\ref{APP:Theorem_R1}.
	\end{IEEEproof}
\end{proposition}

Based on Appendix \ref{APP:Theorem_R1}, we find that the partial derivative of the effective throughput with respect to $R_1$ is monotonically decreasing in terms of $R_1$, as its second partial derivative is less than zero. Thus, the one-dimensional search based on bisection search can be introduced to find the optimal $R_1$.

We note that for a given power allocation we can first determine the value of $R_2$ as per Step 1 and then obtain the value of $R_1$ with the aid of Proposition~\ref{Proposition_R1}. The final hurdle for the optimal design arises from the power allocation. To address this issue, we adopt the one-dimensional numerical search to find the optimal power allocation. In addition, for ensuring the effective throughput target achieved at u$_2$, we next determine the feasible set for power allocation by calculating the lower bound on $P_2$.

\textbf{Step~3}: Determine a strict lower bound on $P_2$.

\begin{lemma}\label{Lemma_R2}
	The strict lower bound on $P_2$, denoted by $P_2^l$, is the unique solution to
	\begin{align}
	R_2^{\ddag}\left(1 - Q\left(f(\gamma_2^l,  R_2^{\ddag})\right)\right) = T_0,
	\end{align}
	where $R_2^{\ddag}$ is satisfies $\mathcal{U}(\gamma_2^l,R_2^{\ddag})=0$, which is defined in \eqref{eq:First_order_T}, and $\gamma_2^l={P_2^lh_2}/({(P - P_2^l)h_2+1})$.
	\begin{IEEEproof}
		The detailed proof is provided in Appendix \ref{APP:Optimal_R2}.
	\end{IEEEproof}
\end{lemma}
For a similar  reason as the previous steps, we cannot derive a closed-form expression for the lower bound on $P_2$. Hence, the one-dimensional search based on bisection search is introduced to obtain the lower bound on $P_2$.

Following Lemma~\ref{Lemma_R2}, we note that the feasible value range of $P_2$ is $P_2^l \leq P_2 \leq P$. This is due to the fact that $\overline{T}_2 \geq T_0$ cannot be guaranteed when $P_2 < P_2^l$ for any possible value of $R_2$ subject to $P_1 + P_2 \leq P$.

\textbf{Step~4}: Determine the optimal power allocation.

Following the aforementioned three steps, the problem given in \eqref{eq:Opt_NOMA} can be simplified to the following optimization problem
\begin{subequations}\label{eq:Opt_NOMA_simplify}
	\begin{align}
	\max_{P_1}~~ &R_1^{\dag}(1-\overline{\epsilon}_1)\label{eq:EffectiveThroughput_OptimalRate}\\
	\st~~ & P_1\in \mathcal{P},
	\end{align}
\end{subequations}
where $\mathcal{P}\triangleq\{p|0\leq p\leq P-P_2^l, P_2=P-p, R_2^{\dag}(1-\overline{\epsilon}_2)=T_0\}$,
which can be solved by the  one-dimensional  line search algorithm over the feasible set of $P_1$. However, we can further draw the following lemma to improve the computational efficiency.
\begin{lemma}\label{Lemma_Unimodal}
	The optimization problem \eqref{eq:Opt_NOMA_simplify} is strictly convex in $P_1$ if 
	\begin{subequations}\label{Cond28}
		\begin{align}
		&\log_2(1+\gamma_1')<R_1^{\dag}\leq \log_2(1+\gamma_1),\\
		&R_2^{\dag}\leq \min\left\lbrace\log_2(1+\gamma_2),~~ \log_2(1+\gamma_2^1)-\frac{2}{Nh_1\ln 2} \right\rbrace.
		\end{align}
	\end{subequations}
	
	\begin{IEEEproof}
		The detailed proof is provided in Appendix~\ref{APP:Unimodal}.
	\end{IEEEproof}
\end{lemma}
It is noted that due to the assumption of large channel disparity between $h_1$ and $h_2$ for fully exploiting the benefit of NOMA, we have that  $\gamma_2^1>\gamma_2$. Furthermore, $\frac{2}{Nh_1\ln 2}$ is negligible since $N\geq 100$ is assumed in the fundamental work~\cite{FBL_2010} for accuracy and a high channel gain is assumed at u$_1$. Thus, the sufficient condition, i.e., $R_2\leq \log_2(1+\gamma_2)-\frac{2}{Nh_1\ln 2}$, can be guaranteed with a high probability based on the fact that $R_2\leq \log(1+\gamma_2)$. 

This lemma shows that the optimal $P_1$ can be found by golden section search instead of line search when $R_1^\dag$ and $R_2^\dag$ satisfy  \eqref{Cond28}.
		

To facilitate the optimal power allocation design, we first choose a feasible value of $P_2$ that guarantees $P_2^l \leq P_2 \leq P$ and then determine $R_2$ for the chosen $P_2$ as per Step~1. After that, $R_1$ is determined as per Step~2 with $P_1 = P - P_2$, since $P_1 + P_2 = P$ is always guaranteed as proved in Lemma~\ref{Lemma_P}. Finally, we can calculate the achieved $\overline{T}_1$ for the chosen $P_2$ and then we repeat the above steps until we find the optimal value of $P_2$ that achieves the maximum value of $\overline{T}_1$, which is denoted by $\overline{T}_1^{\ast}$. Once the optimal value of $P_2$, denoted by $P_2^{\ast}$, is determined, the optimal values of $P_1$, $R_1$, and $R_2$ can be determined accordingly, which are denoted by $P_1^{\ast}$, $R_1^{\ast}$, and $R_2^{\ast}$, respectively.

\section{Design of OMA with a Finite Blocklength}\label{sec:OMA_opt}
In this section, we present the OMA scheme as the benchmark, where the two users are served in different (orthogonal) time slots and hence we have $N_1+N_2= N$.

\subsection{Transmission to Two Users with OMA}

When the AP adopts OMA to serve the two users in orthogonal time slots, the received signal at u$_i$, $i\in\{1,2\}$ is given by 
\begin{align}
y_i = \sqrt{P_{i}}\tilde{h}_i x_i + n_i.
\end{align}
Due to the orthogonal transmissions to u$_1$ and u$_2$, there is no interference at u$_1$ (or u$_2$) caused by $x_2$ (or $x_{1}$). As such, the SNR at u$_i$ of $x_i$ is given by
\begin{align}
\gamma_i=P_ih_i,
\end{align}
where $h_i=\frac{|\tilde{h}_i|^2}{\sigma_i^2}$ denotes the noise normalized channel gain from AP to u$_i$, $i\in\{1,2\}$.

Accordingly, the decoding error probability of $x_i$ at u$_i$ for given $R_i$ is approximated by $\epsilon_i$ given in \eqref{eq:EP}. In addition, the effective decoding error probability $\overline{\epsilon}_i$ is $\epsilon_i$ in OMA, since u$_i$ decodes its own message $x_i$ independently in different time slots. Then, the effective throughput achieved by u$_i$ is given by
\begin{align}\label{T_i_OMA}
\overline{T}_i=\frac{N_i}{N}R_i\left(1-Q(f(\gamma_i,N_i,R_i))\right).
\end{align}

We note that given a time slot allocation, $\overline{T}_1$ is a function of only $P_1$ and $R_1$, while $\overline{T}_2$ is a function of only $P_2$ and $R_2$. This is different from the case in the NOMA scheme, where $\overline{T}_1$ is a function of $P_1$, $R_1$, $P_2$, and $R_2$. The design of OMA is detailed in the following subsection.

\subsection{Optimal Design of OMA}

In the OMA scheme, we have $N_1 + N_2 = N$. As such, for the OMA scheme the optimization problem given in \eqref{eq:Opt} can be rewritten as
\begin{subequations}\label{eq:OptOMA}
	\begin{align}
	\underset{\Delta_o}{\max} ~~~&\overline{T}_1 \label{eq:ObjectiveOMA}\\
	\st~~~&\overline{T}_2\geq T_0,\label{eq:C_T_oma}\\
	&N_1 P_1+ N_2 P_2 \leq NP, \label{eq:C_Energy_OMA}\\
	&N_1+N_2 = N, \label{eq:C_N_OMA}
	\end{align}
\end{subequations}
where $\Delta_o=\{R_1, R_2, P_1, P_2, N_1, N_2\}$ is the variable set that needs to be determined at the AP with OMA transmission. We note that the AP has to determine the time slots allocation on top of optimizing the transmission rates and the power allocation in the OMA scheme.

In order to solve \eqref{eq:OptOMA}, we first clarify that the equalities in \eqref{eq:C_T_oma} and \eqref{eq:C_Energy_OMA} are always guaranteed. This is due to the fact that following \eqref{T_i_OMA} for any given $R_1$, $R_2$, $N_1$, and $N_2$, the effective throughput $T_1$ and $T_2$ are monotonically increasing functions of $P_1$ and $P_2$ respectively. We now briefly outline the steps to solve the optimization problem given in \eqref{eq:OptOMA}.

\textbf{Step~1:} Determine $P_2$ and $R_2$ for a given $N_2$.

For a given time slot allocation (i.e., $N_2$), the optimal value of $P_2$ is the minimum one that guarantees $\overline{T}_2 = T_0$, where the optimal value of $R_2$ is the one that maximizes $\overline{T}_2$ for a given $P_2$. This is due to the fact that $\overline{T}_1$ monotonically decreases with $P_2$ due to $N_1P_1 + N_2P_2 = NP$ and $\overline{T}_1$ is independent of $R_2$.

\textbf{Step~2:} Determine $P_1$ and $R_1$ for given $P_2$ and $N_1$.

Once $P_2$ is determined in Step~1, we have $P_1 =(NP - N_2P_2)/N_1$, since the equality in \eqref{eq:C_Energy_OMA} is always guaranteed. Then, for given $P_1$ and $N_1$, the optimal value of $R_1$ is the one that maximizes $\overline{T}_1$, since $\overline{T}_1$ is independent of $R_2$.

\textbf{Step~3:} Search for the optimal time slot allocation.

Following the aforementioned two steps, we can see that the optimization problem given in \eqref{eq:OptOMA} can be simplified to a one-dimensional numerical search problem in order to determine the optimal values of $N_1$ and $N_2$ subject to $N_1 + N_2 = N$, which can be easily solved. Specifically, we first choose a value of $N_2$ that guarantees $1 \leq N_2 \leq N$ and then determine $P_2$ and $R_2$ as per Step~1. After that, we can determine $P_1$  and $R_1$ as per Step~2. Finally, we can calculate the achieved $\overline{T}_1$ for the chosen value of $N_2$ and then we repeat the above steps until we find the optimal value of $N_2$ that achieves the maximum $\overline{T}_1$, denoted by $\overline{T}_1^{\ast}$. Once the optimal value of $N_2$, denoted by $N_2^{\ast}$, is determined, the optimal values of $N_1$, $P_1$, $P_2$, $R_1$, and $R_2$ can be determined accordingly, which are denoted by $N_1^{\ast}$, $P_1^{\ast}$, $P_2^{\ast}$, $R_1^{\ast}$, and $R_2^{\ast}$, respectively.

\section{Numerical Results}\label{sec:simulation}

\begin{figure}[t!]
	\begin{center}
		\includegraphics[width=3.5in]{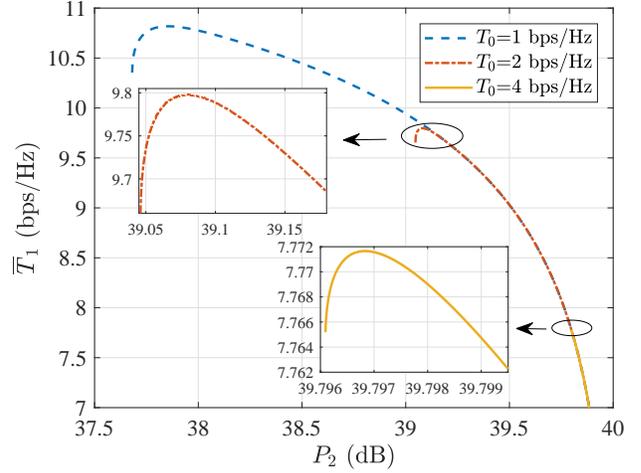}\vspace{-0mm}
		\caption{Effective throughput $\overline{T}_1$ achieved by the NOMA scheme versus $P_2$ with different values of $T_0$, where $|\tilde{h}_1|=0.8$, $|\tilde{h}_2|=0.2$, $\overline{\gamma}=40\dB$, and $N=100$.}
		\label{fig:T1_P2}
	\end{center}\vspace{0mm}
\end{figure}

In this section, we present numerical results to examine the performance of the proposed NOMA scheme with the OMA scheme as the benchmark by considering the finite blocklength. We assume that the AP and the users are located in a 30m $\times$ 80m factory. The distances between AP and the two users are set to be $d_1=20$m and $d_2=60$m, respectively. Unless otherwise stated, we set the noise power at each user to unit, i.e., $\sigma_{1}^{2}=\sigma_{2}^{2}=1$. We also define the average transmit SNR as $\overline{\gamma} = {\overline{P}}/{\sigma_i^2}$. Besides, we set $\overline{T}_1$ to zero if the optimization problem in \eqref{eq:Opt} is infeasible, i.e., if the constraint $\overline{T}_2 \geq T_0$ cannot be guaranteed, to incorporate the penalty of failure. 

\subsection{Numerical Results Based on Fixed Channel Gains}

In order to provide the insight into the relationships between the objective function and the parameters of interest, throughout this subsection, the channel gains of the two users are set to be fixed.

In Fig.~\ref{fig:T1_P2}, we plot the effective throughput of u$_1$ achieved by the NOMA scheme, i.e., $\overline{T}_1$, versus $P_2$, while other parameters (e.g., $P_1$, $R_1$, $R_2$) are optimized accordingly. In this figure, we first observe that the lower bound on $P_2$, i.e., $P_2^l$, increases with $T_0$. When $P_2$ is smaller than $P_2^l$, u$_2$ cannot achieve the target effective throughput $T_0$, which leads to that $\overline{T}_1$ is set to be zero when $P_2 < P_2^l$. We also observe that the optimal $P_2$ that maximizes $\overline{T}_1$, i.e., $P_2^{\ast}$, is not $P_2^l$. This is caused by the high-order terms of the effective error probability in \eqref{eq:Average_EP1}. This observation also verifies our analysis presented in Section~\ref{subsec:Optimal_Design} that $\overline{T}_1$ may not be a monotonically increasing function of $P_1 = P- P_2$, which is significantly different from the case with an infinite blocklength. The observation $P_2^{\ast} \geq P_2^l$ indicates that in the NOMA scheme with a finite blocklength, more power is allocated to u$_2$ relative to in NOMA with an infinite blocklength, which enables us to adopt a smaller $R_2$ in order to reduce the outage probability of SIC in the NOMA scheme with a finite blocklength (SIC can be guaranteed in NOMA with an infinite blocklength). To confirm this, we examine the optimal value of $R_2$ in the following figure.

\begin{figure}[t!]
	\begin{center}
		\includegraphics[width=3.5in]{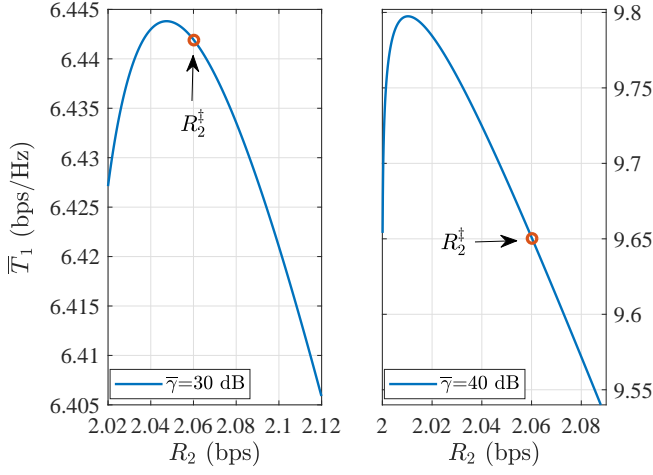}
		\caption{Effective throughput $\overline{T}_1$ achieved by the NOMA scheme versus $R_2$ with different values of $\overline{\gamma}$, where $T_0=2$ bps/Hz, $N=200$, $|\tilde{h}_1|=0.8$, and $|\tilde{h}_2|=0.1$.}
		\label{fig:T1_R2}
	\end{center}
\end{figure}

In Fig.~\ref{fig:T1_R2}, we plot the effective throughput achieved at u$_1$, i.e., $\overline{T}_1$, versus $R_2$ with different values of the average transmit SNR, while other system parameters are optimized accordingly. In this figure, $\overline{T}_1$ achieved by $R_2^{\ddag}$, which is corresponding to the lower bound on $P_2$ (i.e., $P_2^l$), is marked in red circles. As expected, we first observe that the optimal $R_2$ that maximizes $\overline{T}_1$, i.e., $R_2^{\ast}$, is lower than  $R_2^{\dag}$. This confirms the fact detailed in the previous paragraph, which is due to the fact that the outage probability of SIC increases with $R_2$ and is explained in our Lemma~\ref{Lemma_T2}. Together with Fig.~\ref{fig:T1_P2}, we find that more power is allocated while a smaller transmission rate is set for u$_2$ in order to maximize $\overline{T}_1$ subject to $\overline{T}_2 \geq T_0$  in the NOMA scheme with a finite blocklength, relative to in the one with an infinite blocklength. This is mainly due to the fact that the outage probability of SIC monotonically decreases with $P_2$ and monotonically increases with $R_2$. In Fig.~\ref{fig:T1_R2}, we further observe that the difference between $R_2^{\ast}$ and $R_2^{\ddag}$ increases with the average transmit SNR $\overline{\gamma}$, since we can use more power to counteract the impact of the reduction in $R_2$ when $\overline{\gamma}$ is high. This also verifies that the outage probability of SIC decreases with $\overline{\gamma}$, which indicates that we prefer to guarantee SIC in the NOMA scheme with a finite blocklength when there is enough transmit power.


\begin{figure}[t!]
	\begin{center}
		\includegraphics[width=3.5in]{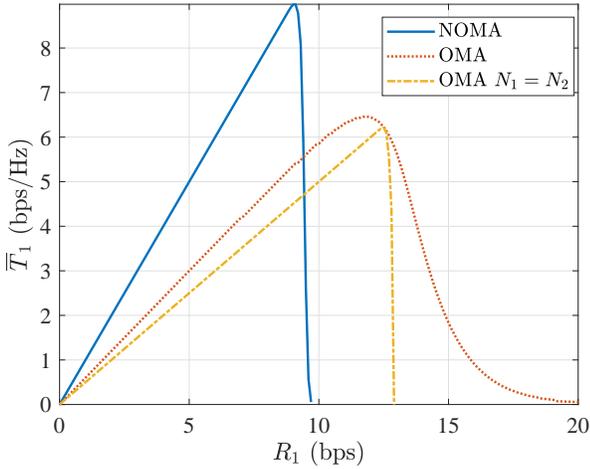}
		\caption{Effective throughput $\overline{T}_1$ achieved by the NOMA, OMA and OMA  with a fixed time slot allocation schemes versus $R_1$, where $T_0=3$ bps/Hz, $\overline{\gamma}=40\dB$, $N=200$, $|\tilde{h}_1|=0.8$, and $|\tilde{h}_2|=0.1$.}
		\label{fig:NOMA_OMA_R1}
	\end{center}
\end{figure}

\begin{figure}[t!]
	\centering
		\includegraphics[width=3.5in]{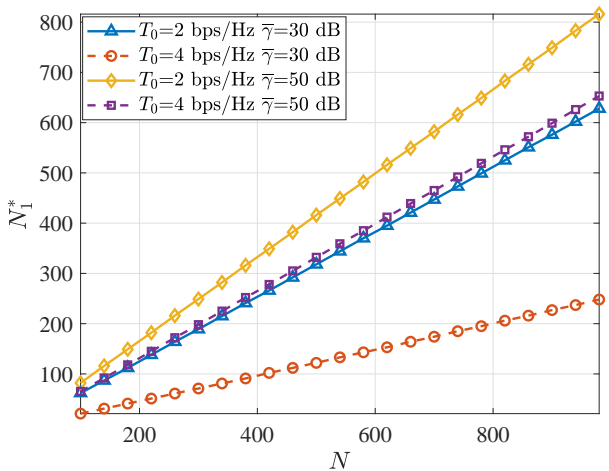}
		\caption{The optimal time slots allocated to u$_1$ versus $N$ in the OMA scheme with different values of $T_0$ and $\overline{\gamma}$, where $|\tilde{h}_1|=0.8$ and $|\tilde{h}_2|=0.1$.}
		\label{fig:N1_N}
\end{figure}

In Fig.~\ref{fig:NOMA_OMA_R1}, we plot the effective throughput achieved at u$_1$, i.e., $\overline{T}_1$, versus $R_1$. Specifically, we compare the proposed NOMA scheme with the optimal OMA scheme and fixed time slot allocation OMA scheme, where $N_1=N_2=N/2$, which is denoted by ``OMA $N_1=N_2$''. It is noted that for any given $R_1$, the power allocation and channel coding rate design for u$_2$ in NOMA scheme, ``OMA $N_1=N_2$'' scheme, and the additional time slot allocation in OMA scheme are optimized. In this figure, we first observe that there exists a unique optimal value of $R_1$ that maximizes $\overline{T}_1$ in both the NOMA and OMA schemes, which verifies our Proposition~\ref{Proposition_R1}. Second, we observe that for the optimal $R_1$, NOMA significantly outperforms OMA in terms of achieving a higher $\overline{T}_1$, which demonstrates the advantage of NOMA in short-packet communications to improve the spectral efficiency. However, the optimal value of transmission rate $R_1^{\ast}$ in NOMA is lower than that in OMA. This is due to the co-channel interference incurred by NOMA. As such, the throughput advantage offered by NOMA is achieved by the overloading spectrum but with a lower transmission rate.  Furthermore, we observe that the maximum $\overline{T}_1$ achieved by the OMA scheme with $N_1=N_2$ is lower than that achieved by OMA. This demonstrates the necessity of optimizing the time slot allocation in the OMA scheme, which will be examined in Fig. \ref{fig:N1_N}. Finally, we observe that as $R_1$ increases, the effective throughput $\overline{T}_1$ first increases and then decreases. In fact, there is a non-trivial trade-off between the effective throughput and the transmission rate.  When $R_1$ is small, the effective error probability of u$_1$ is significantly small and has almost negligible impact on the effective throughput of u$_1$. As such, the throughput performance is limited by the transmission rate.  However, when $R_1$ is beyond a certain value, the corresponding error probability increases exponentially. Thus, the effective error probability plays a dominant role and results in a sharp decrease in the effective throughput.

\begin{figure}[!t]
	\begin{center}
		\includegraphics[width=3.5in]{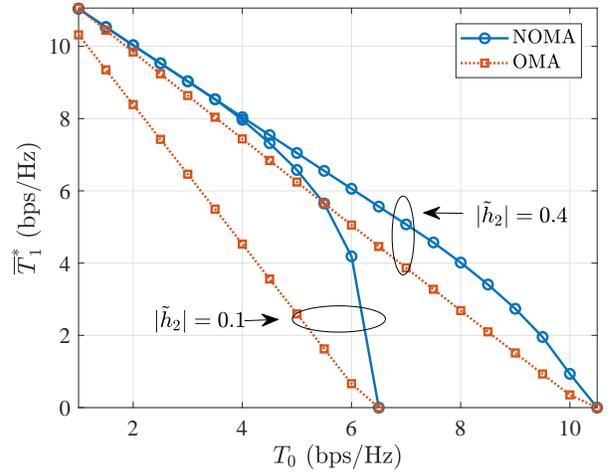}
		\caption{The maximum effective throughput $\overline{T}_1^{\ast}$ achieved by the NOMA and OMA schemes versus $T_0$ with different values of $h_2$, where $\overline{\gamma}=40\dB$, $|\tilde{h}_1|=0.8$, and $N=200$.}\label{fig:T1_T0}
	\end{center}	
\end{figure}

In Fig.~\ref{fig:T1_T0}, we plot the maximum effective throughput $\overline{T}_1^{\ast}$ achieved by the proposed NOMA scheme and OMA scheme versus $T_0$, with different channel gains from the AP to u$_2$, i.e., $|\tilde{h}_2|$. As expected, in this figure we first observe that $\overline{T}_1^{\ast}$ monotonically decreases with $T_0$. This is due to the fact that the equality in the constraint~\eqref{eq:C_T_NOMA} is always active, as proved in Lemma~\ref{Lemma_T2}, and more resources need to be allocated to u$_2$ as $T_0$ increases. This observation demonstrates the trade-off between the achieved effective throughput at the two users. We also observe that $\overline{T}_1^{\ast}$ achieved by NOMA is higher than that achieved by OMA and the performance gap is maximized in the regime where $\overline{T}_1^{\ast}$ is close to $T_0$. This demonstrates that the advantage of the NOMA scheme becomes more dominant when the fairness (in terms of the achieved effective throughput) between the two users is desirable, i.e., when the two uses require similar effective throughput. This is different from the infinite blocklength scenario, where the performance gap increases with $T_0$, e.g., \cite{Chen_2017_TSP,Xu_2017_JSAC}, since the optimization of time slot allocation is introduced in the proposed finite blocklength OMA scheme. In addition, we observe that the performance gap between NOMA and OMA increases as the disparity between $h_1$ and $h_2$ increases, which is also observed in the NOMA scheme with an infinite blocklength \cite{Ding_CM_2017} and confirms that NOMA is more preferred when the channel gains of the two users to the AP are significantly different. The last two observations indicate that with a finite blocklength the NOMA scheme is more desirable than the OMA scheme in the scenario where the two uses have significantly different channel gains but require similar effective throughput.

In Fig.~\ref{fig:N1_N}, we plot the optimal number of time slots allocated to u$_1$, i.e., $N_1^{\ast}$, in the OMA scheme versus the total number of time slots $N$, which allows us to examine the optimal time slot allocation strategy in the OMA scheme. In this figure, we first observe that $N_1^{\ast}$  approximately linearly increases with $N$. This demonstrates that time slot allocation plays an important role in maximizing $\overline{T}_1$ since time slots are precious resources in short-packet communications. We also observe that $N_1^{\ast}$ increases with $\overline{\gamma}$, which means that the optimal number of time slots allocated to u$_2$, i.e., $N_2^{\ast} = N - N_1^{\ast}$, decreases with $\overline{\gamma}$. This indicates that when the transmit power increases, more power is allocated to u$_2$  in stead of the time slots to guarantee $\overline{T}_2 = T_0$, which confirms that time slots are more precious in the scenario with a finite blocklength. Finally, as expected we observe that $N_1^{\ast}$ decreases with $T_0$, since more time slots need to be allocated to u$_2$ in order to guarantee $\overline{T}_2 = T_0$ as $T_0$ increases.

\begin{figure}[t!]
	\begin{center}
		\includegraphics[width=3.5in]{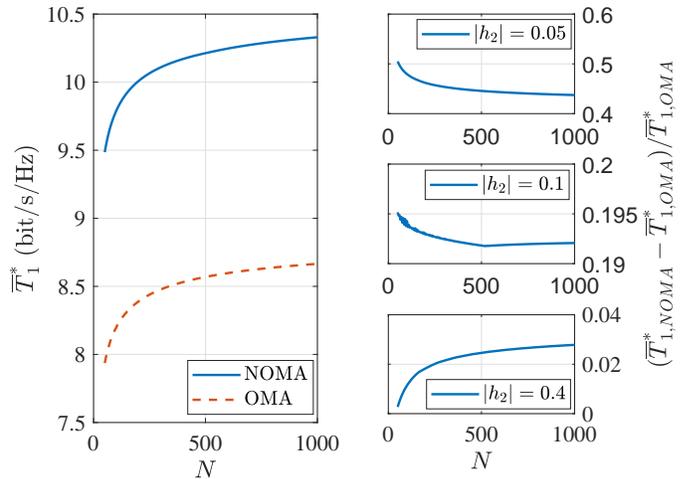}
		\caption{The maximum effective throughput $\overline{T}_1^{\ast}$  achieved by the NOMA and OMA schemes versus the blocklength $N$, where $\overline{\gamma}=40\dB$, $|\tilde{h}_1|=0.8$, and $|\tilde{h}_2|=0.1$.}
		\label{fig:T1_N}
	\end{center}
\end{figure}

\begin{figure}[!t]
	\centering
		\includegraphics[width=3.5in]{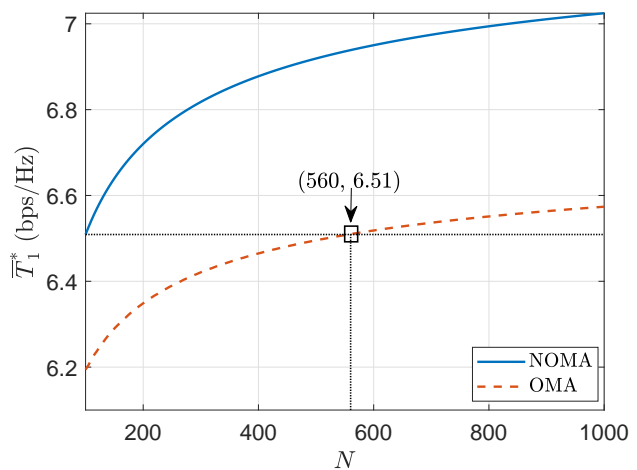}
		\caption{The maximum effective throughput $\overline{T}_1^{\ast}$ achieved by the NOMA and OMA schemes versus $N$, where $\overline{\gamma}=30\dB$, $|\tilde{h}_1|=0.8$, $|\tilde{h}_2|=0.4$, and $T_0=2$ bps/Hz.}\label{fig:T1_N_h2}
\end{figure}

In Fig.~\ref{fig:T1_N} and Fig.~\ref{fig:T1_N_h2},we examine the impact of the blocklength $N$ on the performance of the NOMA and OMA schemes. Specifically, in Fig.~\ref{fig:T1_N}(a) and Fig.~\ref{fig:T1_N_h2} we plot the maximum effective throughput $\overline{T}_1^{\ast}$ achieved by the NOMA and OMA schemes versus the blocklength $N$ for different channel gain disparities, while in Fig.~\ref{fig:T1_N}(b) we plot the performance gap between these two schemes versus $N$. In Fig.~\ref{fig:T1_N}(a) and Fig.~\ref{fig:T1_N_h2}, we observe that the proposed NOMA scheme significantly outperforms the OMA scheme regardless of $N$. We see from Fig.~\ref{fig:T1_N}(a) that when the channel gain disparity between the two users is large, the NOMA scheme can significantly reduce the communication delay in short-packet communications. For example, when $\overline{T}_1 = 9.5$ bps/Hz, we find that the NOMA scheme can achieve the desired $\overline{T}_1$ with $N = 50$, while the OMA scheme cannot achieve $\overline{T}_1$ even with an infinite $N$. Moreover, by decreasing the channel gain disparity between the two users, as illustrated in Fig.~\ref{fig:T1_N_h2}, we find that NOMA still has a performance advantage over OMA. For example, to achieve $\overline{T}_1 = 6.51$ bps/Hz, NOMA requires $100$ blocklength, i.e., $N=100$, while OMA requires $560$ blocklength, i.e., $N=560$. This demonstrates that NOMA significantly lowers the latency relative to OMA for achieving the same effective throughput target.  We note that the short and finite blocklength also has a negative impact on the NOMA scheme, since the SIC cannot be always guaranteed with a finite blocklength. As such, the blocklength $N$ does not have a unique impact on the performance gap between these two schemes, i.e., this performance gap may increase or decrease with $N$ for different system settings (e.g., different values of $|\tilde{h}_2|$ for a fixed $|\tilde{h}_1|$). This is confirmed by our Fig.~\ref{fig:T1_N}(b).



\subsection{Numerical Results Based on Random Channel Gains}

In this subsection, we model the channels as $\tilde{h}_i=d_i^{-\alpha}\bar{h}_i, ~i\in\{1,2\}$, where $\bar{h}_i$ represents the Rayleigh fading coefficients from the AP to u$_i$, and $\alpha$ is the path loss exponent, which is set to be $2$. The entries in $\bar{h}_i$ are modeled as CSCG random variables with zero mean and unit variance. The simulation results are averaged over 10,000 independent channel realizations.

\begin{figure}[!t]
	\begin{center}
		\includegraphics[width=3.5in]{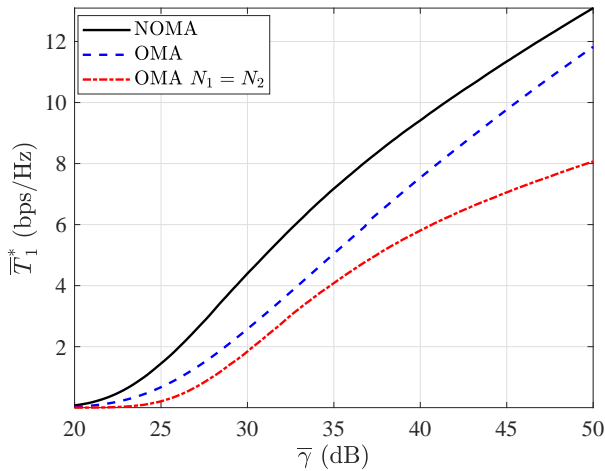}
		\caption{The maximum effective throughput $\overline{T}_1^{\ast}$ achieved by the NOMA and OMA schemes versus $\overline{\gamma}$, where $T_0=2$ bps/Hz, and $N=200$.}\label{fig:T1_SNR}
	\end{center}
\end{figure}

In Fig.~\ref{fig:T1_SNR}, we plot the maximum $\overline{T}_1$, i.e., $\overline{T}_1^{\ast}$, achieved by the NOMA, OMA, and fixed time slot allocation OMA schemes versus  $\overline{\gamma}$. In this figure, we first observe that the proposed NOMA scheme always outperforms the OMA scheme regardless the value of $\overline{\gamma}$. We also observe that the performance gap between NOMA and OMA first increases with $\overline{\gamma}$ and then decreases. This is different from the comparison result of these two schemes in the scenario with an infinite blocklength, where this performance gap always increases with $\overline{\gamma}$. This is due to the fact that time slot allocation is optimized in the OMA scheme with a finite blocklength, while the time allocation is not considered in the OMA scheme with an infinite blocklength OMA. This is confirmed by another observation in this figure, which is that the performance gap between the NOMA scheme and the OMA scheme with $N_1 = N_2$ always increases with $\overline{\gamma}$.

\section{Conclusion}\label{sec:conclusion}
This work introduced NOMA in short-packet communications for the IoT to achieve low latency. An optimization problem was addressed to maximize the effective throughput of the user with a higher channel gain while ensuring that the other user achieved a certain level of the effective throughput. To facilitate the optimal design of transmission rates and power allocation for the two NOMA users, the analysis and insights on the constraints were provided and an optimal solution was proposed. With OMA serving as a benchmark, the analytical and numerical examinations demonstrated that NOMA outperforms OMA by achieving a much higher effective throughput. This also indicates that NOMA significantly reduces the latency in short-packet communications for achieving the same effective throughput. We further found that NOMA offers a profound fairness advantage over OMA, as the performance gap between NOMA and OMA becomes more prominent as the effective throughputs at the two users become more comparable. The two-user single-antenna scenario investigated in this work serves as a foundation for more general scenarios. One promising future direction is to jointly design user clustering and transmission strategy in a massive-user scenario. Another one is to introduce multiple or even massive antennas into the system for further improving the spectral efficiency.

\appendices
\section{Proof of Proposition~\ref{lemmaEP_P}}\label{App:EP_Gamma}
We now prove that the decoding error probability $\epsilon_i$ given in \eqref{eq:EP} is a monotonically decreasing function of the corresponding SNR or SINR, i.e., $\gamma_i$. To this end, we derive the partial derivative of $\epsilon_i$ with respect to $\gamma_i$  as
\begin{align}\label{eq:EP_gamma}
\frac{\partial \epsilon_i}{\partial \gamma_i}=-\frac{1}{\sqrt{2\pi}}e^{-\frac{f^2\left(\gamma_i, N_i, R_i\right)}{2}}\frac{\partial f\left(\gamma_i, N_i, R_i\right)}{\partial \gamma_i},
\end{align}
where $\frac{\partial f\left(\gamma_i, N_i, R_i\right)}{\partial \gamma_i}$ is the partial derivative of $f\left(\gamma_i, N_i, R_i\right)$ with respect to $\gamma_i$ and is given by
\begin{align}\label{eq:f'gamma}
\frac{\partial f\left(\gamma_i, N_i, R_i\right)}{\partial \gamma_i}=\sqrt{N_i}\frac{1-\ln2\frac{\log_2(1+\gamma_i)-R_i}{(1+\gamma_i)^2-1}}{\sqrt{(1+\gamma_i)^2-1}}.
\end{align}
However, it is not obvious to know whether $\frac{\partial f\left(\gamma_i, N_i, R_i\right)}{\partial \gamma_i} > 0$ or not. To address this issue, we first define a function by $\mathcal{G}(x)=\frac{\log_2x}{x^2-1}$. We next determine the value range of  $\mathcal{G}(x)$ for $x\geq 1$ since $1+\gamma_i >1$ in \eqref{eq:f'gamma}.
To this end, we derive the first derivative of $\mathcal{G}(x)$ with respect to $x$ as
\begin{align}\label{Gx_dev}
\mathcal{G}'(x)=\frac{g(x)}{\left(x^2-1\right)^2},
\end{align}
where $g(x)$ is given by $g(x) = \frac{1}{\ln 2}\left(x-\frac{1}{x}\right)-2x\log_2 x$.
We note that the sign of $\mathcal{G}'(x)$ is the same as $g(x)$ due to $\left(x^2-1\right)^2 > 0$ when $x >1$. The first derivative of $g(x)$ with respect to $x$ is given by $g'(x)=-\frac{1}{\ln2}\left(1-\frac{1}{x^2}\right)-2\log_2x$. We note that $g'(x) < 0$ for $x >1$, which means that  $g(x)$ is a decreasing function of $x$ when $x > 1$ and thus $g(x)< g(1)=0$.
Then, as per \eqref{Gx_dev} we know that $\mathcal{G}'(x)< 0$ when $x > 1$, which leads to the fact that $\mathcal{G}(x)$ is a decreasing function of $x$ when $x\geq 1$. As per the L'Hospital's rule, we have $\mathcal{G}(x)$ approaches to $\frac{1}{2\ln2}$ and $0$ as $x$ approaches to $1$ and the positive infinity, respectively, which means that $0 < \mathcal{G}(x)< \frac{1}{2\ln 2}$ when $x >1$. As such, noting $R_i \geq 0$ and $\gamma_i > 0$ we have
\begin{align}\label{key_proof}
1-\ln2\frac{\log_2(1+\gamma_i)-R_i}{(1+\gamma_i)^2-1} &\geq  1-\ln2\frac{\log_2(1+\gamma_i)}{(1+\gamma_i)^2-1}\notag\\
&=1-\ln2 \mathcal{G}(\gamma_i+1)\notag \\
&> 1 - \ln 2 \frac{1}{2\ln 2}= \frac{1}{2}>0.
\end{align}
Then, following \eqref{eq:f'gamma} and \eqref{key_proof} we have $\frac{\partial f\left(\gamma_i, N_i, R_i\right)}{\partial \gamma_i}>0$, which leads to $\frac{\partial \epsilon_i}{\partial \gamma_i}< 0$ as per \eqref{eq:EP_gamma}. Therefore, we can conclude that $\epsilon_i$ is a monotonically decreasing function of $\gamma_i$. We note that $\epsilon_i$ can be any one of $\epsilon_1$, $\epsilon_1'$, $\epsilon_2^1$, and $\epsilon_2$, since they are functions of $\gamma_1$, $\gamma_1'$, $\gamma_2^1$, and $\gamma_2$, respectively, and these functions are all defined by \eqref{eq:EP}.

\section{Proof of Lemma~\ref{Lemma_P}}\label{App:Power}

We now prove Lemma~\ref{Lemma_P} by contradiction. We first suppose that the current optimal power allocation $P_1^{\dag}$ and $P_2^{\dag}$, satisfying $P_1^{\dag}+P_2^{\dag}<P$, can achieve the maximum value of $\overline{T}_1$, which is denoted by $T_1^{\dag}$, while guaranteeing the constraint $\overline{T}_2 \geq T_0$. We then increase $P_1^{\dag}$ and $P_2^{\dag}$ by multiplying a common scalar, $\alpha={P}/({P_1^{\dag}+P_2^{\dag}})$, to obtain a new power allocation $P_1^{\ddag} = \alpha P_1^{\dag}$ and $P_2^{\ddag} = \alpha P_2^{\dag}$, satisfying $P_1^{\ddag}+P_2^{\ddag}=P$. We note that $P_1^{\ddag} > P_1^{\dag}$ and $P_2^{\ddag} > P_2^{\dag}$ due to $\alpha > 1$.

Following \eqref{eq:SINR2} and noting $\alpha > 1$, we have
\begin{align}
\gamma_2^{\ddag}=\frac{P_2^{\ddag}h_2}{P_1^{\ddag}h_2+1} &= \frac{P_2^{\dag}h_2}{P_1^{\dag}h_2+\frac{1}{\alpha}}>\frac{P_2^{\dag}h_2}{P_1^{\dag}h_2+{1}} = \gamma_2^{\dag}.
\end{align}
As such, we can conclude that $\gamma_2$ increases as $P_1^{\dag}$ and $P_2^{\dag}$ increase to $P_1^{\ddag}$ and $P_2^{\ddag}$, respectively. Then, following similar proofs we can see that $\gamma_{2}^{1}$, $\gamma_1$, and $\gamma'_{1}$ all increase when $P_1^{\dag}$ increases to $P_1^{\ddag}$ and $P_2^{\dag}$ increases to $P_2^{\ddag}$, as per \eqref{eq:SINR_SIC},~\eqref{eq:SNR1}, and \eqref{eq:SINR1}, respectively. This leads to the fact that the decoding error probabilities $\epsilon_2$, $\epsilon_{2}^{1}$, $\epsilon_1$, and $\epsilon'_{1}$ all decrease based on Proposition~\ref{lemmaEP_P}.

Following \eqref{eq:T2}, we note that $\overline{T}_2$ increases as $\epsilon_2$ decreases and thus the constraint $\overline{T}_2 \geq T_0$ can be still guaranteed by the new power allocation, i.e., $P_1^{\ddag}$ and $P_2^{\ddag}$. Based on \eqref{eq:Average_EP1}, we know that $\overline{\epsilon}_1$ is a monotonically increasing function of $\epsilon_1$ and  $\epsilon'_{1}$ due to $0 \leq \epsilon_{2}^{1} \leq 1$. Then, we examine the monotonicity of $\overline{\epsilon}_1$ with respect to $\epsilon_2^1$. To this end, following \eqref{eq:Average_EP1}, the partial derivative of $\overline{\epsilon}_1$ with respect to $\epsilon_{2}^{1}$ is derived as $\frac{\partial \overline{\epsilon}_1}{\partial \epsilon_2^1} = \epsilon'_{1} - \epsilon_1$. Based on \eqref{eq:SNR1} and \eqref{eq:SINR1}, we note that $\epsilon'_{1} > \epsilon_1$ and thus ${\partial \overline{\epsilon}_1}/{\partial \epsilon_2^1} >0$, which leads to the fact that $\overline{\epsilon}_1$ monotonically increases with $\epsilon_{2}^{1}$ as well.

As $\overline{\epsilon}_1$ is a monotonically increasing function of $\epsilon_1$,  $\epsilon'_{1}$, and $\epsilon_{2}^{1}$, $\overline{\epsilon}_1$ decreases as $P_1^{\dag}$ and $P_2^{\dag}$ increase to $P_1^{\ddag}$ and $P_2^{\ddag}$, respectively. It follows that $\overline{T}_1$ increases from $T_1^{\dag}$ to $T_1^{\ddag}$, i.e., $T_1^{\dag} < T_1^{\ddag}$, as per \eqref{eq:Average_T1}. This contradicts to the claim of optimality that $T_1^{\dag}$ is the maximum value of $\overline{T}_1$ achieved by $P_1^{\dag}$ and $P_2^{\dag}$. Therefore, we conclude that $P_1+P_2=P$ is always guaranteed in the optimal solution to the optimization problem given in \eqref{eq:Opt_NOMA}.

\section{Proof of Lemma~\ref{Lemma_T2_R2}}\label{APP:T2_R2}

To examine the monotonicity and concavity of $\mathcal{T}(R_2)$ in terms of $R_2$, we derive the first and second derivatives of $\mathcal{T}(R_2)$ with respect to $R_2$ in the followings.

The first derivative of $\mathcal{T}\left(R_{2}\right)$ with respect to $R_2$ is derived as
\begin{align}\label{eq:f'R2}
\mathcal{T}'\left(R_{2}\right)=1\!-\!Q\left(f(\gamma_2, R_2)\right)-\frac{R_{2}b}{\sqrt{2\pi}}e^{-\frac{f^2(\gamma_2, R_2)}{2}},
\end{align}
where $b=\frac{\sqrt{N}\ln 2}{\sqrt{1-(1+\gamma_2)^{-2}}}$. Notably, the sign of $\mathcal{T}'\left(R_{2}\right)$ is not always negative nor positive. Hence, $\mathcal{T}(R_2)$ does not monotonically increase nor decrease with $R_2$. 

Then, following \eqref{eq:f'R2}, the second derivative of $\mathcal{T}\left(R_{2}\right)$ with respect to $R_2$ is given by
\begin{align}\label{eq:f''R2}
\mathcal{T}''\left(R_{2}\right)=\frac{-2b}{\sqrt{2\pi}}e^{-\frac{f^2(\gamma_2,R_2)}{2}}-\frac{R_2f(\gamma_2,R_2)b^2}{\sqrt{2\pi}}e^{-\frac{f^2(\gamma_2,R_2)}{2}}.
\end{align}
Noting $b\geq 0$ and $f(\gamma_2,R_2)\geq 0$, we can conclude that $\mathcal{T}''\left(R_{2}\right) \leq 0$, which indicates that $\overline{T}_2$ is a concave function of $R_2$.

\section{Proof of Proposition~\ref{Proposition_R1}}\label{APP:Theorem_R1}

Following \eqref{eq:Average_T1}, we define the effective throughput $\overline{T}_1$ in terms of $R_1$ by
\begin{align}\label{eq:f(R1)}
\mathcal{T}\left(R_1\right)&\triangleq R_1\left(1-Q\left(f\left(\gamma_1, R_1\right)\right)\right)(1-\epsilon_2^1)\notag\\
&+R_1\left(1-Q\left(f\left(\gamma_1', R_1\right)\right)\right)\epsilon_2^1.
\end{align}
The optimal value of $R_1$ that maximizes $\mathcal{T}\left(R_{1}\right)$ is tackled by examining the monotonicity and concavity of  $\mathcal{T}\left(R_{1}\right)$ with respect to $R_1$. To this end, the first and second derivatives of $\mathcal{T}\left(R_{1}\right)$ with respect to $R_1$ are derived in the following.


Based on the differentiation of a definite integral with respect to the input argument~\cite{Table_2007}, the first derivative of $\mathcal{T}\left(R_{1}\right)$ with respect to $R_1$ is derived as
\begin{align}\label{eq:f'R1}
&\mathcal{T}'\left(R_{1}\right)=\notag\\
&\begin{cases}
(1\!-\!\epsilon_2^1)\mathcal{U}(\gamma_1,R_1)\!+\!\epsilon_2^1\mathcal{U}(\gamma_1',R_1),~\text{if}~R_1\!\leq\! \log_2(1\!+\!\gamma_1'),\\
(1-\epsilon_2^1)\mathcal{U}(\gamma_1,R_1),~\text{if}~\log_2(1+\gamma_1')<R_1\leq \log_2(1+\gamma_1),
\end{cases}
\end{align}
where
\begin{align}\notag
\mathcal{U}(x,R_1)=1-Q\left(f(x, R_1)\right)+\frac{R_1\frac{\partial f(x,R_1)}{\partial R_1}}{\sqrt{2\pi}}
e^{-\frac{f^2(x, R_1)}{2}}, 
\end{align}
and
\begin{align}\label{proofT3_2}
\frac{\partial f(x,R_1)}{\partial R_1}=-\frac{\sqrt{N}\ln 2}{\sqrt{1-(1+x)^{-2}}},~x \in \{\gamma_1, \gamma_1'\},
\end{align}
which is the partial derivative of $f(x, R_1)$ with respect to $R_1$. 

Following \eqref{eq:f'R1}, the second derivative of $\mathcal{T}\left(R_{1}\right)$ with respect to $R_1$ is derived as
\begin{align}\label{eq:f''(R1)}
&\mathcal{T}''\left(R_{1}\right)=\notag\\
&\begin{cases}(1\!-\!\epsilon_2^1)\mathcal{V}(\gamma_1,R_1)\!+\!\epsilon_2^1\mathcal{V}(\gamma_1',R_1),~\text{if}~R_1\!\leq\! \log_2(1+\gamma_1'),\\
(1-\epsilon_2^1)\mathcal{V}(\gamma_1,R_1),~\text{if}~\log_2(1+\gamma_1')<R_1\leq \log_2(1+\gamma_1),
\end{cases}
\end{align}
where, for $x\in \{\gamma_1,\gamma_1'\}$, $\mathcal{V}(x,R_1)$ is defined by
\begin{align}\notag
\mathcal{V}(x,R_1)=&\sqrt{\frac{2}{\pi}}\frac{\partial f(x,R_1)}{\partial R_1}e^{-\frac{f^2(x,  R_1)}{2}}-\notag\\
&\frac{R_1f(x,R_1)\left(\frac{\partial f(x,R_1)}{\partial R_1}\right)^2}{\sqrt{2\pi}}e^{-\frac{f^2(x,  R_1)}{2}}.
\end{align}

Following \eqref{proofT3_2}, we note that $\frac{\partial f(\gamma_1,R_1)}{\partial R_1}\leq 0$ and $\frac{\partial f(\gamma_1',R_1)}{\partial R_1}\leq 0$. We also note that $f(\gamma_1', R_1) \geq 0$ for $R_1\!\leq\! \log_2(1+\gamma_1')$ and $f(\gamma_1, R_1)\geq 0$ for $R_1\leq \log_2(1+\gamma_1)$. As such, we can conclude that $\mathcal{T}''(R_1)\leq 0$ holds for the two value ranges of $R_1$ given in \eqref{eq:f''(R1)}.This means that $\mathcal{T}\left(R_1\right)$ is concave with respect to $R_1$ in the reasonable value range of $R_1$, i.e.,  $R_1\leq \log_2(1+\gamma_1)$. This indicates that the optimal value of $R_1$ that maximizes the effective throughput $\overline{T}_1$ can be achieved by setting $\mathcal{T}'\left(R_{1}\right)=0$, which completes the proof of Proposition~\ref{Proposition_R1}.

\section{Proof of Lemma~\ref{Lemma_R2}}\label{APP:Optimal_R2}


In order to prove this theorem, we first examine the monotonicity of $P_2$ with respect to $R_2$. Based on Lemma~\ref{Lemma_T2}, $R_2$ and $P_2$ need to guarantee the equality in constraint \eqref{eq:C_T_NOMA}, i.e., $\overline{T}_2 = T_0$, for maximizing $\overline{T}_1$. For the sake of clarity, we first define 
\begin{align}\label{defined_equation}
F(R_2,P_2)=R_2(1-\epsilon_2)-T_0=0.
\end{align}
Following \eqref{defined_equation} and applying the implicit function theorem \cite{Implicit_2003}, the first derivative of $P_2$ with respect to $R_2$ is given by
\begin{align}
\frac{\partial P_2}{\partial R_2}=-\frac{\partial F/\partial R_2}{\partial F/\partial P_2}=-\frac{\partial \overline{T}_2/\partial R_2}{\partial \overline{T}_2/\partial P_2}.
\end{align}

We next examine the value ranges of ${\partial \overline{T}_2/\partial R_2}$ and ${\partial \overline{T}_2/\partial P_2}$ in order to determine the value range of ${\partial P_2}/{\partial R_2}$.
As per the definition of $\overline{T}_2$ given in \eqref{eq:T2}, the partial derivative of $\overline{T}_2$ with respect to $P_2$ is derived as
\begin{align}\label{T2P2}
\frac{\partial \overline{T}_2}{\partial P_2}=-\frac{\partial \epsilon_2}{\partial \gamma_2}\frac{\partial \gamma_2}{\partial P_2}.
\end{align}
Following Proposition~\ref{lemmaEP_P}, we note that $\epsilon_2$ is a decreasing function of $\gamma_2$. Noting $\gamma_2$ increases with $P_2$ as per \eqref{eq:SINR2}, we conclude that $\overline{T}_2$ monotonically increases with $P_2$, i.e.,   $\frac{\partial \overline{T}_2}{\partial P_2}\geq 0$.

As for $\partial \overline{T}_2/\partial R_2$, based on Appendix~\ref{APP:T2_R2}, we find that $\overline{T}_2$ is a concave function of $R_2$. This indicates that the value of $R_2$ which maximizes $\overline{T}_2$ is unique. We denote this value by $R_2^{\ddag}$ and obtain $\mathcal{T}'(R_2^{\ddag})=0$. Then, we have  ${\partial \overline{T}_2}/{\partial R_2}> 0$ when $R_2\leq R_2^{\ddag}$, whereas ${\partial \overline{T}_2}/{\partial R_2}< 0$ when $R_2 > R_2^{\ddag}$. In addition, the value of $P_2$ that guarantees $\overline{T}_2=T_0$ and associates with $R_2^{\ddag}$ is denoted by $P_2^l$. Following \eqref{eq:f''R2}, \eqref{defined_equation}, and \eqref{T2P2},  we note that with the constraint $\overline{T}_2 = T_0$, $P_2$ decreases with $R_2$ when $R_2 \leq R_2^{\ddag}$ and $P_2$ increases with $R_2$ when $R_2 > R_2^{\ddag}$. As such, we can conclude that $P_2^l$ is the lower bound on $P_2$, which completes the proof of Lemma~\ref{Lemma_R2}.
\vspace{-0mm}
\section{Proof of Lemma~\ref{Lemma_Unimodal}}\label{APP:Unimodal}

Based on \eqref{eq:Average_T1}, the effective throughput of u$_1$ with $\log_2(1+\gamma')\leq R_1 \leq \log_2(1+\gamma)$ is $\overline{T}_1=R_1(1-\epsilon_1)(1-\epsilon_2^1)$. To explore the concavity of $\overline{T}_1$ with respect to $P_1$, the first-order and second-order partial derivatives are derived as
\begin{align}
\frac{\partial \overline{T}_1}{\partial P_1}=-R_1(1-\epsilon_1)\frac{\partial \epsilon_2^1}{\partial P_1}-R_1(1-\epsilon_2^1)\frac{\partial \epsilon_1}{\partial P_1},
\end{align}
and
\begin{align}\label{eq:T_2d_P1}
\frac{\partial^2 \overline{T}_1}{\partial P_1^2} = -R_1\left(-2\frac{\partial \epsilon_1}{\partial P_1}\frac{\partial \epsilon_2^1}{\partial P_1}+(1-\epsilon_2^1)\frac{\partial^2 \epsilon_1}{\partial P_1^2}+(1-\epsilon_1)\frac{\partial^2 \epsilon_2^1}{\partial P_1^2}\right).
\end{align}

Notably, $\frac{\partial \epsilon_2^1}{\partial P_1}\geq 0$ and $\frac{\partial \epsilon_1}{\partial P_1}\leq 0$ are proved in Appendix~\ref{App:EP_Gamma}. This indicates that the first term in the bracket in \eqref{eq:T_2d_P1} is positive.  However, the sign of the second-order partial derivative of $\epsilon_i$ with respect to $P_1$ in the second and third terms in \eqref{eq:T_2d_P1} is not clear yet. To address this issue, the second-oder partial derivative of $\epsilon_i$ with respect to $P_1$ is given by
\begin{align}\label{eq:EP_2d_P1}
\frac{\partial^2 \epsilon_i}{\partial P_1^2}=\frac{\partial^2\epsilon_i}{\partial \gamma_i^2}\left(\frac{\partial \gamma_i}{\partial P_1}\right)^2+\frac{\partial \epsilon_i}{\partial \gamma_i}\frac{\partial^2 \gamma_i}{\partial P_1^2}.
\end{align}
Recall that $\frac{\partial \epsilon_i}{\partial \gamma_i}\leq 0$, which is derived in Appendix~\ref{App:EP_Gamma}, and  $\left(\frac{\partial \gamma_i}{\partial P_1}\right)^2$ must be positive due to the quadratic form. Then, to deal with the sign of $\frac{\partial^2 \epsilon_i}{\partial P_1^2}$,  we first derive the expression for the second-order partial derivative of $\epsilon_i$ with respect to its corresponding $\gamma_i$, i.e., $\frac{\partial^2\epsilon_i}{\partial \gamma_i^2}$, and then derive the expression for $\frac{\partial^2 \gamma_i}{\partial P_1^2}$ in the followings. 

Based on the partial derivative of $\epsilon_i$ with respect to $\gamma_i$ in \eqref{eq:EP_gamma}, the second-order partial derivative of $\epsilon_i$ with respect to $\gamma_i$ is given by
\begin{align}\label{eq:EP_2d_Gamma}
\frac{\partial^2\epsilon_i}{\partial \gamma_i^2}=-\frac{1}{\sqrt{2\pi}}e^{-\frac{f^2\left(\gamma_i,  R_i\right)}{2}}\left(-2f\left(\gamma_i, R_i\right)\left(\frac{\partial f\left(\gamma_i, R_i\right)}{\partial \gamma_i}\right)^2+\frac{\partial^2 f\left(\gamma_i, R_i\right)}{\partial \gamma_i^2}\right).
\end{align}
Then, based on the partial derivative of $f\left(\gamma_i, R_i\right)$ with respect to $\gamma_i$ in \eqref{eq:f'gamma}, the expression for the second-order partial derivative of $f\left(\gamma_i, R_i\right)$ with respect to $\gamma_i$ is given by 
\begin{align}\label{eq:f_2d_gamma}
\frac{\partial^2 f\left(\gamma_i, R_i\right)}{\partial \gamma_i^2}=&a
\left(-\frac{1}{1+\gamma_i}-(1+\gamma_i)\right)\left((1+\gamma_i)^2-1\right)+\notag\\
&3a(1+\gamma_i)\left(\log_2(1+\gamma_i)-R_i\right)\ln 
2.
\end{align}
where $a=\frac{\sqrt{N}}{\left((1+\gamma_i)^2-1\right)^{\frac{5}{2}}}$. 
To tackle the sign of the term after the product sign, we define a function by $\mathcal{K}(x)$,  which is given by 
\begin{align}\label{eq:k_x}
\mathcal{K}(x)=\left(-\frac{1}{x}-x\right)(x^2-1)+3\ln 2 (\log_2x-R_i)x,
\end{align}
where $x\geq 1$. 
The sign of $\mathcal{K}(x)$ is dealt with its first derivative and second derivative with respect to $x$.  To this end, the first derivative of $\mathcal{K}(x)$ with respect to $x$ is derived as
\begin{align}\label{eq:k_1d_x}
\mathcal{K}'(x)=-3x^2-\frac{1}{x}+3\ln 2(\log_2x-R_i)+3,
\end{align}
and the second derivative of $\mathcal{K}(x)$ with respect to $x$ is derived as
\begin{align}\label{eq:k_2d_x}
\mathcal{K}''(x)=-6x+\frac{2}{x^3}+\frac{3}{x}=\frac{-6\left(x^2-\frac{1}{4}\right)^2+\frac{19}{8}}{x^3}.
\end{align}
Based on the expression for $\mathcal{K}''(x)$ in \eqref{eq:k_2d_x}, we have that $\mathcal{K}''(x)$ monotonically decreases with $x$ when $x\geq \frac{1}{2}$. Thus, $\mathcal{K}''(x)\leq \mathcal{K}''(1)=-1$, when $x\geq 1$. This indicates that $\mathcal{K}'(x)$ monotonically decreases with $x$ when $x\geq 1$. Accordingly, we have $\mathcal{K}'(x)\leq \mathcal{K}'(1)=-1-3\ln 2 R_i\leq 0$. This results in a monotonically decreasing trend of $\mathcal{K}(x)$ with respect to $x$ when $x\geq 1$. Therefore, $\mathcal{K}(x)\leq \mathcal{K}(1)=-3\ln 2 R_i\leq 0$. Then, substituting $\gamma_i+1$ for $x$ in $\mathcal{K}(x)$ in \eqref{eq:k_x}, we have the term after the product sign in \eqref{eq:f_2d_gamma} is negative. As such, we have that $\frac{\partial^2 f\left(\gamma_i, R_i\right)}{\partial \gamma_i^2}\leq 0$.  

Since $f(\gamma_i, R_i)\geq 0$, $\left(\frac{\partial f\left(\gamma_i, R_i\right)}{\partial \gamma_i}\right)^2\geq 0$, and $\frac{\partial^2 f\left(\gamma_i, R_i\right)}{\partial \gamma_i^2}\leq 0$, based on the expression for $\frac{\partial^2\epsilon_i}{\partial \gamma_i^2}$ in \eqref{eq:EP_2d_Gamma}, we have that $	\frac{\partial^2\epsilon_i}{\partial \gamma_i^2}\geq 0$.

We then deal with the first and second derivatives of $\gamma_i$ with respect to $P_1$.  Based on the definition of $\gamma_2^1$ and $\gamma_1$ in \eqref{eq:SINR_SIC} and \eqref{eq:SNR1}, respectively, the partial derivatives of $\gamma_2^1$ and $\gamma_1$ with respect to $P_1$ are given by
\begin{align}\label{eq:gamma_1d_p1}
\frac{\partial \gamma_2^1}{\partial P_1}=\frac{-h_1-Ph_1^2}{(P_1h_1+1)^2}\leq 0, ~~~\frac{\partial \gamma_1}{\partial P_1}=h_1\geq 0.
\end{align}
Following \eqref{eq:gamma_1d_p1}, the second-order partial derivatives of $\gamma_2^1$ and $\gamma_1$ with respect to $P_1$ are given by
\begin{align}\label{eq:gamma_2d_p1}
\frac{\partial^2 \gamma_2^1}{\partial P_1^2}=\frac{2(h_1+Ph_1^2)}{(P_1h_1+1)^3}\geq 0, ~~~\frac{\partial^2 \gamma_1}{\partial P_1^2}=0.
\end{align}

Based on \eqref{eq:EP_2d_Gamma}, \eqref{eq:gamma_1d_p1},  and \eqref{eq:gamma_2d_p1}, the sign of the second-order partial derivative of $\epsilon_1$ with respect to $P_1$ in \eqref{eq:EP_2d_P1} can be determined, i.e., $\frac{\partial^2 \epsilon_1}{\partial P_1^2}\geq 0$. But that of $\epsilon_2^1$ with respect to $P_1$ is not clear. This is due to the fact that the first term in \eqref{eq:EP_2d_P1} for $\epsilon_2^1$ is positive while the second term is negative. This brings difficulty to deal with the sign. To further address this issue, we then derive the expression for $\frac{\partial^2 \epsilon_2^1}{\partial P_1^2}$, which is given by
\begin{align}\label{eq:EP2_2d_P1}
\frac{\partial^2 \epsilon_2^1}{\partial P_1^2}=&\frac{\partial^2\epsilon_2^1}{\partial (\gamma_2^1)^2}\left(\frac{\partial \gamma_2^1}{\partial P_1}\right)^2+\frac{\partial \epsilon_2^1}{\partial \gamma_2^1}\frac{\partial^2 \gamma_2^1}{\partial P_1^2}\notag\\
=&-\frac{2}{\sqrt{2\pi}}e^{-\frac{f^2\left(\gamma_2^1,  R_i\right)}{2}}\frac{\partial f\left(\gamma_2^1, R_i\right)}{\partial \gamma_2^1}\frac{h_1+Ph_1^2}{(P_1h_1+1)^3}\notag\\
&\times\left(-f\left(\gamma_2^1, R_i\right)h_1\frac{\partial f\left(\gamma_2^1, R_i\right)}{\partial \gamma_2^1}(1+\gamma_2^1) +1\right)\notag\\
&-\frac{2}{\sqrt{2\pi}}e^{-\frac{f^2\left(\gamma_2^1,  R_i\right)}{2}}\frac{\partial^2 f\left(\gamma_2^1, R_i\right)}{\partial (\gamma_2^1)^2}\frac{(h_1+Ph_1^2)^2}{(P_1h_1+1)^4}.
\end{align}
It is noted that the second term in \eqref{eq:EP2_2d_P1} is positive. The difficulty is to tackle the sign of the first term. To address this issue, we deal with the function in the bracket in the first term of \eqref{eq:EP2_2d_P1}. Its expression is given by
\begin{subequations}
	\begin{align}
	&-f\left(\gamma_2^1, R_i\right)h_1\frac{\partial f\left(\gamma_2^1, R_i\right)}{\partial \gamma_2^1}(1+\gamma_2^1) +1\label{eq:f1}\\
	&=\frac{-N\ln 2  \left(\log_2(1+\gamma_2^1)-R_2\right)\left(1-\ln 2 \frac{\log_2(1+\gamma_2^1)-R_2}{(1+\gamma_2^1)^2-1}\right)}{1-\frac{1}{(1+\gamma_2^1)^2}}h_1+1\label{eq:f2}\\
	&< -\frac{N h_1\ln 2 }{2} \frac{\log_2(1+\gamma_2^1)-R_2}{1-\frac{1}{(1+\gamma_2^1)^2}}+1\label{eq:f3}\\
	&< -\frac{N h_1\ln 2 }{2}(\log_2(1+\gamma_2^1)-R_2)+1. \label{eq:f4}
	\end{align}
\end{subequations}
The first inequality in \eqref{eq:f3} is due to the fact that $ 1-\ln 2 \frac{\log_2(1+\gamma_2^1)-R_2}{(1+\gamma_2^1)^2-1}> \frac{1}{2}$ based on \eqref{key_proof}. And the second inequality is owning to $1-\frac{1}{(1+\gamma_2^1)^2}< 1$. 

Thus, if $R_2\leq \log_2(1+\gamma_2^1)-\frac{2}{Nh_1\ln 2}$ satisfies, the second-order derivative of $\epsilon_2^1$ with respect to $P_1$, i.e., $\frac{\partial^2 \epsilon_2^1}{\partial P_1^2}$, is larger than zero. 

Recall the expression in \eqref{eq:T_2d_P1}, the first term in that is positive, the second term is positive since $\frac{\partial^2 \epsilon_1}{\partial P_1^2}\geq 0$ based on \eqref{eq:EP_2d_P1}, \eqref{eq:EP_2d_Gamma}, \eqref{eq:gamma_1d_p1}, and \eqref{eq:gamma_2d_p1}, and the third term is positive when $R_2\leq \log_2(1+\gamma_2^1)-\frac{2}{Nh_1\ln 2}$, due to $\frac{\partial^2 \epsilon_2^1}{\partial P_1^2}\geq 0$. As a consequence, we have $\frac{\partial^2 \overline{T}_1}{\partial P_1^2}\leq 0$. This indicates that $\overline{T}_1$ is strictly concave with respect to $P_1$ with a sufficient condition, i.e., 
\begin{align}
R_2\leq \min \left\lbrace \log_2(1+\gamma_2), \log_2(1+\gamma_2^1)-\frac{2}{Nh_1\ln 2} \right\rbrace.
\end{align}


\renewcommand\refname{References}~
\bibliographystyle{IEEEtran}
{\footnotesize\bibliography{IEEEabrv,Reference}}
\end{document}